\documentclass[aps,prx,reprint,superscriptaddress,amsmath,amssymb,floatfix,nofootinbib]{revtex4-2}

\usepackage[dvipsnames]{xcolor}

\definecolor{teal}{HTML}{4633FF}
\usepackage{graphicx}%
\usepackage{dcolumn}%
\usepackage{bm}%
\usepackage[colorlinks=true]{hyperref}%
\hypersetup{
  linkcolor  = teal,
  citecolor  = Green!90!Black,
  urlcolor   = Purple
}
\usepackage{cleveref}
\usepackage{makecell}
\usepackage{csquotes}
\usepackage{placeins}
\usepackage{enumitem}
\setlist[itemize]{noitemsep}
\input{Qcircuit}
\newcommand{\minus}{\scalebox{0.5}[1.0]{$-$}}
\newcommand{\Gi}{\mathcal{X}}
\newcommand{\es}{\hspace{0.1 ex}\underline{\hspace{1 ex}}\hspace{0.1 ex}}
\newcommand{\os}{\hspace{0.1 ex}\resizebox{1 ex}{!}{$\ast$}\hspace{0.1 ex}} 
\newcommand{\GiTerm}{\Gi\Gi\Gi\Gi}
\newcommand{\PiTerm}{\Pi\Pi\Pi\Pi}
\newcommand{\FredkinTrick}{Subspace Gating}
\newcommand{\UCDBT}{Uniformly-Controlled Qudit Rotations}%
\newcommand{\CCDBT}{Control-Restricted Qudit Rotations}%
\newcommand{\DBT}{\CCDBT}%
\newcommand{\qtplaq}{\hat{\square}^{(1)}}
\newcommand{\qdplaq}{\hat{\square}}
\newcommand{\ctrlseq}{\bm{C}}
\newcommand{\ccnot}[3]{\mathcal{R}^{#2,#1}_#3(\ctrlseq,\bm{\theta})}
\newcommand{\manccnot}[3]{\tilde{\mathcal{R}}^{#2,#1}_#3(\ctrlseq,\bm{\theta})}
\newcommand{\ucnot}[3]{\mathcal{R}^{#2,#1}_#3(\bm{\theta})}
\renewcommand{\arraystretch}{1.4}

\newcommand{\sumpi}{\left( \sum_\Pi \right)}
\newcommand{\onexxxxunitary}{\exp(-i \tau \sumpi \GiTerm)}
\newcommand{\uqtplaq}{e^{-i \tau \qtplaq}}

\newcommand{\uqdeplaq}{e^{-i\tau \hat{\square}^{(\Lambda_j)}}}
\newcommand\scalemath[2]{\scalebox{#1}{\mbox{\ensuremath{\displaystyle #2}}}}

\creflabelformat{equation}{#2\textup{#1}#3}

\begin{document}

\title{Non-Abelian dynamics on a cube: improving quantum compilation through qudit-based simulations}

\author{Jacky Jiang}
\email{jacky.j@alumni.ubc.ca}
\affiliation{Department of Electrical and Computer Engineering, The University of British Columbia, Vancouver, BC, Canada}
\author{Natalie Klco}
\email{natalie.klco@duke.edu}
\affiliation{Duke Quantum Center and Department of Physics, Duke University, Durham, NC 27708}
\author{Olivia Di Matteo}
\email{olivia.dimatteo@ubc.ca}
\affiliation{Department of Electrical and Computer Engineering, The University of British Columbia, Vancouver, BC, Canada}
\affiliation{Stewart Blusson Quantum Matter Institute, The University of British Columbia, Vancouver, BC, Canada}

\date{\today}
\preprint{APS/123-QED}

\begin{abstract}
\noindent
Recent developments in mapping lattice gauge theories relevant to the Standard Model onto digital quantum computers identify scalable paths with well-defined quantum compilation challenges toward the continuum.
As an entry point to these challenges, we address the simulation of SU(2) lattice gauge theory.
Using qudit registers to encode the digitized gauge field, we provide quantum resource estimates, in terms of elementary qudit gates, for arbitrarily high local gauge field truncations.
We then demonstrate an end-to-end simulation of real-time, qutrit-digitized SU(2) dynamics on a cube. 
Through optimizing the simulation, we improved circuit decompositions for uniformly-controlled qudit rotations, an algorithmic primitive for general applications of quantum computing.
The decompositions also apply to mixed-dimensional qudit systems, which we found advantageous for compiling lattice gauge theory simulations.
Furthermore, we parallelize the evolution of opposite faces in anticipation of similar opportunities arising in three-dimensional lattice volumes.
This work details an ambitious executable for future qudit hardware and attests to the value of codesign strategies between lattice gauge theory simulation and quantum compilation.
\end{abstract}

\maketitle

\section{Introduction}
\label{sec:intro}
\noindent
Gauge field theories describe three of the four fundamental forces of our universe.  
Lattice gauge theory (LGT) provides a systematic regularization of such field theories amenable to computational approaches.
In LGT, the field is discretized onto a spatial lattice~\cite{Wilson:1974,Kogut:1975} in which bosonic (fermionic) content of the theory resides on links (sites).  
As a non-perturbative framework, lattice techniques are well-suited to address the complex emergent phenomena ubiquitous in gauge theories. 

The expanse and density of a lattice reflects the accuracy of the associated discretized field representation.
Reducing the lattice spacing (raising the ultraviolet truncation) and increasing the volume (lowering the infrared truncation) to systematically improve resolution of physical length scales recovers the continuum behavior of gauge theories.
However, this requires large Hilbert spaces to accurately capture dynamical, highly entangled processes in nuclear and high-energy physics.
As such, many experimentally-relevant observables remain intractable, even in the era of exascale computing~\cite{Joo:2019byq,Osti:2022}.

Quantum computers are expected to efficiently simulate LGTs~\cite{Byrnes:2005qx,Banuls:2019bmf,Aidelsburger:2021mia,Bauer:2023,Bauer:2023qgm}.
Many representations of LGT on quantum computers are actively being developed~\cite{Byrnes:2005qx,Zohar:2012xf,Banerjee:2012xg,Zohar:2013zla,Zohar:2014qma,Zohar:2015hwa,Zohar:2018cwb,Raychowdhury:2019iki,Klco:2019evd,Lamm:2019bik,Davoudi:2020yln,Buser:2020cvn,Ji:2020kjk,Kreshchuk:2020dla,Bauer:2021gup,Ciavarella:2021nmj,Wiese:2021djl,
Haase:2020kaj,Kadam:2022ipf,Davoudi:2022xmb,Pardo:2022hrp,Liu:2023lsr,Zache:2023dko,Ciavarella:2023mfc,DAndrea:2023qnr,Ciavarella:2024fzw,Grabowska:2024emw,Kavaki:2024ijd,Halimeh:2024bth,Ciavarella:2025bsg,Balaji:2025afl}.
The computational resource consequences of each representation---e.g., circuit width and depth, ratio of physical to unphysical Hilbert space and sensitivity to noise, compilation efficiency and required classical preprocessing, and rapidity of convergence to the continuum---vary dramatically.
While innovative representations continue to arise, persistent features of scalable strategies involve mapping quantum registers to local quantum numbers of the gauge field and retaining a non-zero gauge-variant subspace within the computational Hilbert space of the quantum device.

The latticized gauge field is represented by an infinite-dimensional Hilbert space local to each link,
making qudits a natural candidate for LGT simulations as they encode gauge fields at a higher truncation than qubits.
Qudit-based quantum computing hardware has also advanced rapidly in recent years~\cite{Lu:2020jux,Cervera-Lierta:2021nhp,Ringbauer:2021lhi,Morvan:2021qju,Hrmo:2022bvo,Goss:2022bqd,Nguyen:2023svc,Seifert:2023ous,Low:2023dlg},
incentivizing further exploration of qudit-based quantum simulations~\cite{Senko:2015akp,Blok:2020may,Ciavarella:2021nmj,Kurkcuoglu:2021dnw,Gonzalez-Cuadra:2022hxt,Zache:2023cfj,Popov:2023xft,Halimeh:2023lid,Vezvaee:2024vuj,Illa:2023scc,Meth:2023wzd,Turro:2024shh,Illa:2024kmf,Calajo:2024qrc,Kurkcuoglu:2024cfv}.
Moreover, qudits show promise for scaling universal quantum computing in general applications~\cite{Gottesman:1998se,Campbell:2014dok,Inada:2021sgc,Cowtan:2022csx,Brock:2024vkc,Muthukrishnan:2000,Nikolaeva:2021rhq,Lanyon:2009,Li:2013,Liu:2020,Baker:2020,Litteken:2022,Saha:2020rob,Majumdar:2024sms,Lysaght:2024omz,Wang:2020,Fischer:2022,Deller:2023,Bottrill:2023lyt,Zahidy:2024wux,Popp:2024isu},
placing importance on understanding qudit circuit synthesis techniques and associated costs ~\cite{Zhou:2003,Bullock:2005,Brennen:2005,Khan:2006,Nakajima:2009,Di:2013,Luo:2014cmp,Luo:2014,Di:2015,Zi:2023,Jiang:2023,Mato:2024,Yang:2025wcj,Prakash:2018,Yeh:2022,vandeWetering:2022doq,Kalra:2024gnc,Glaudell:2024tvz,Kalra:2023llu,Gustafson:2025buk}.

In this work, we design and compile a qudit-based quantum simulation of LGT on a cube.
We utilize the representation developed in Refs.~\cite{Klco:2019evd,Ciavarella:2021nmj,ARahman:2021}, i.e., a link-extended local basis storing only the total angular momentum at each link.
This basis is scalable, has reduced proportion of unphysical subspace, requires fewer quantum registers, and offers a clear path for explicit quantum compilation into elementary gates.
By engaging in the compilation, we mutually advance qudit-based simulation of LGT and fundamental qudit compilation techniques with the following contributions:
\begin{itemize}
    \item improved circuit decompositions for qudit state preparation and unitary synthesis,
    \item optimized synthesis of the arbitrary-truncation (qudit) plaquette operator evolution,
    \item full specification of an executable for qutrit-digitized SU(2) gauge field simulation on a cube.
\end{itemize}
Our first contribution benefits qudit-based quantum computing in general.
For instance, the improved decomposition of {\UCDBT} (\Cref{subsec:ucdbt}) has direct applications to qudit state preparation~\cite{Bullock:2005,Brennen:2005} and unitary synthesis~\cite{Bullock:2005,Brennen:2005,Khan:2006,Nakajima:2009,Di:2013,Di:2015,Yang:2025wcj}.
Furthermore, the decomposition of {\CCDBT} (\Cref{subsec:ccdbt}) is useful in mixed-dimensional qudit settings. In particular, we find  value in incorporating auxiliary qudits whose dimension scales only with lattice connectivity rather than gauge field truncation.

Time-evolution of the plaquette operator comprises the dominant cost of LGT simulations.
As our second contribution, we efficiently synthesize the unitary evolution of the plaquette operator and provide precise quantum resource estimates in terms of elementary qudit gates.
The synthesis procedures apply to any gauge field truncation, providing an explicit recipe for simulating LGT on quantum hardware.

Our third contribution serves as verification of the decompositions, and as an ambitious executable for future qutrit hardware~\cite{qlgt-code}.
Our use of a cube highlights parallelization opportunities that arise at the
operator level when time evolution involves shared registers in higher-dimensional lattices.

Below, we begin with an explanation of the target LGT simulation and derive properties of the qudit plaquette operator in \Cref{sec:su2cube}.
The improved qudit decomposition techniques are presented in \Cref{sec:decompositions}.
These techniques are used to efficiently synthesize the qutrit cube simulation in \Cref{sec:qutrit-cube-sim}, and to report explicit resource counts for unitary evolution of the qudit plaquette operator  in \Cref{sec:qudit-decomp}.
We discuss future works in \Cref{sec:results}.
In optimizing the target LGT simulation, we glean insights into quantum architectural design and compilation methods for general applications of qudit-based quantum computing,
which is reminiscent of the role LGT played in driving HPC development~\cite{Marinari:1986,Boyle:2004,Christ:2011}.
To that end, this work is written from both LGT and quantum compilation perspectives, separately and together, with the goal of facilitating ongoing codesign.

\section{Pure SU(2) Gauge Field on a Cube}
\label{sec:su2cube}
\noindent
Demonstrating techniques that extend naturally to larger lattices, we simulate SU(2) lattice gauge theory on a cube with no matter fields 
\begin{figure}
    \includegraphics[width=\columnwidth]{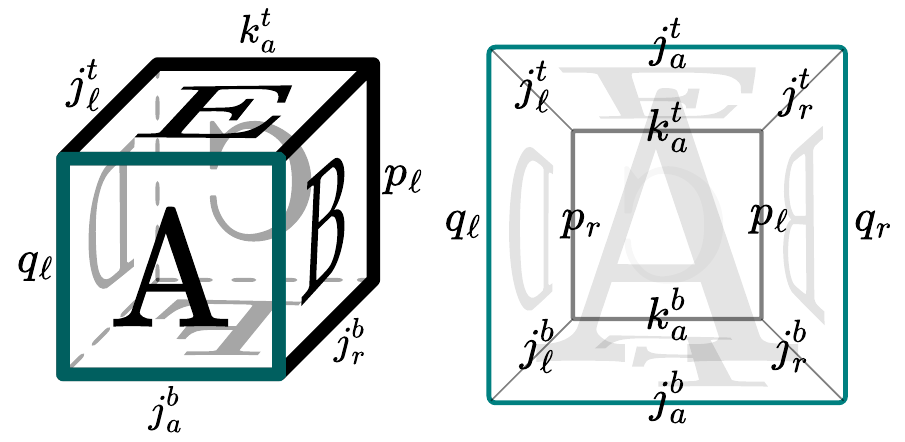}
    \caption{
        We use single qudits to encode the gauge field at each link.
        While the developed decomposition techniques apply to Trotterizations for many lattice configurations of plaquettes, the simulation demonstrated is a cube.
        Applying the plaquette operator to face A evolves the (teal-colored) plaquette links $\ket{q_\ell\ j^b_a\ q_r\ j^t_a}$ controlled on the neighboring links $\ket{j^t_\ell\ j^b_\ell\ j^b_r\ j^t_r}$.
    }
    \label{fig:system} 
\end{figure}
(\Cref{fig:system}). We use the angular momentum basis and remove alignment variables by incorporating part of the local gauge symmetry \cite{Klco:2019evd}.
The remaining Hilbert space is represented by a single quantum register on each gauge link.
Registers are structured to represent a one-dimensional tower of states labeled by total angular momentum $j$ spanning half integers from 0 to the truncation $\Lambda_j$.
Absent matter (both static and dynamical), physical lattice configurations in this basis satisfy the triangle inequalities of angular momentum addition at each vertex, and continue to do so throughout the unitary dynamics.
In our simulation, we use $d$-level quantum systems (qudits) to encode the gauge field using one qudit at each link.

To capture the non-Abelian dynamics, we use the Hamiltonian of Yang-Mills lattice gauge theory \cite{Kogut:1975}, which is composed of an electric and magnetic term,
\begin{align}
    \hat{H} &= \hat{H}_E + \sum_{\text{faces}}\hat{H}_\square \nonumber \\
            &= \frac{g^2}{2} \sum_{\text{links}}{\hat{E}^2} - \frac{1}{2g^2}\sum_{\text{faces}}\left( \hat{\square} + \hat{\square}^\dagger\right).
    \label{eq:yang-mills}
\end{align}
For the SU(2) gauge theory, the plaquette operator of the magnetic term is Hermitian, $\hat{\Box} = \hat{\Box}^\dagger$,
and the local gauge-invariant Casimir operator is $\hat{E}^2=\sum_{j} j(j+1)\Pi_{2j}$, where $\Pi_i=\ket{i}\bra{i}$ is the projection operator in the computational basis.

\subsection{Truncated Plaquette Operator}
\label{subsec:plaq-op}
\noindent
The matrix elements of the plaquette operator $\qdplaq$ generate transitions between states in our chosen basis of total angular momentum quantum number per link.
The Hilbert space of the operator comprises a physical subspace, $\mathcal{P}$ (where states contract to form a gauge singlet at each vertex), and an unphysical subspace, $\bar{\mathcal{P}}$.
The plaquette operator $\qdplaq$ can be expressed as a direct sum, $\hat{\Box} = \qdplaq_{\mathcal{P}} \oplus \qdplaq_{\bar{\mathcal{P}}}$, reflecting the separation of these subspaces throughout time evolution.

Initializing our simulation in $\mathcal{P}$ grants the freedom to assign arbitrary (Hermiticity-preserving) values to matrix elements of $\hat{\Box}_{\bar{\mathcal{P}}}$. 
This freedom can be leveraged to introduce a gauge-variant completion (GVC) that simplifies the eventual quantum circuit implementation~\cite{Klco:2019evd}.
We choose a GVC for $\qdplaq$ with the form
\begin{equation}
    \hat{\Box} \stackrel{GVC}{=} \sum_{ijklpqrs} \phi_{ijklpqrs} \Pi_i \Pi_j\Pi_k \Pi_l \Gi_{p p'} \Gi_{q q'} \Gi_{r r'} \Gi_{ss'},
    \label{eq:plaquetteGVC}
\end{equation}
where the Hermitian operators $\Gi_{xy} = \ket{x}\bra{y} + \ket{y}\bra{x}$ are restricted to adjacent subspaces $\Gi_{x x^\prime} = \Gi_{x,x+1}$ due to the ladder connectivity of irreducible representations (irreps) of SU(2) dynamics generated by \Cref{eq:yang-mills}.
The physical transition matrix elements $\phi$ are obtained from \Cref{eq:angle-formula} \cite{Klco:2019evd,ARahman:2021}.

Insight into the structure of $\qdplaq$ can be obtained by visualizing the flux line configurations of the three links at a vertex of the cube,
\begin{figure}
    \includegraphics[width=\columnwidth]{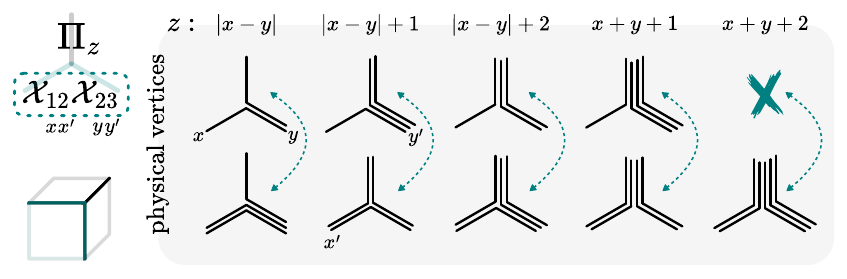}%
    \caption{\label{fig:vertex-transfers} 
        Graphical representation of how $\Gi_{x x^\prime}\Gi_{y y^\prime}$ affects the flux lines at a vertex via example for $\Gi_{12}\Gi_{23}$.
        The number of flux lines is $2j$, representing the link's total angular momentum.
        The operator $\Gi_{x x^\prime}\Gi_{y y^\prime}$ transitions the plaquette link flux lines between two physical configurations indicated by the arrows.
        The control values corresponding to physical configurations span from $|x-y|$ to $x+y+2$.
        However, applying $\Gi_{x x^\prime}\Gi_{y y^\prime}$ to the flux configuration with $x+y+2$ results in an unphysical state,
        and thus $\Pi_{x+y+2}$ is not paired with $\Gi_{x x^\prime}\Gi_{y y^\prime}$ in $\qdplaq$.
    }
\end{figure}
as depicted in \Cref{fig:vertex-transfers}.
To connect all valid SU(2) irreps, the operator $\qdplaq$ must contain a complete set of $(d-1)^4$ distinct $\Gi_{p p'} \Gi_{q q'} \Gi_{r r'} \Gi_{ss'}$, since there are $(d-1)$ choices for each $\Gi$ in $\GiTerm$ (indices suppressed for clarity).

Gauge-invariance is preserved by conditional application of  $\Gi_{p p'} \Gi_{q q'} \Gi_{r r'} \Gi_{ss'}$~\cite{Ciavarella:2022zhe}.
Applying $\qdplaq$ to plaquette A in \Cref{fig:system} would apply $\Pi_i\Pi_j\Pi_k\Pi_l$ to the control links $\ket{j^t_\ell j^b_\ell j^b_r j^t_r}$,
thus controlling the application of $\Gi_{p p'} \Gi_{q q'} \Gi_{r r'} \Gi_{ss'}$ to the plaquette links $\ket{q_\ell j^b_a q_r j^t_a}$ with control word $ijkl$.
The control value $i$ of a valid gauge-invariant transition at the vertex (\Cref{fig:vertex-transfers}) satisfies $i \in F(x,y)$, 
\begin{align}
    F(x,y) = \{|x-y|, |x-y|+1, \cdots, x+y+1\},
    \label{eq:control-set}
\end{align}
where $0 \leq i \leq d-1$ since the gauge field is digitized to a $d$-dimensional qudit.

To ensure the system is physical, a further constraint is imposed by Gauss's law via a Gaussian surface that cuts through all four control links of $\qdplaq$.
The set of projectors $\mathcal{C}_{pqrs}$ controlling $\Gi_{pp^\prime}\Gi_{qq^\prime}\Gi_{rr^\prime}\Gi_{ss^\prime}$ becomes
\begin{align}
    \mathcal{C}_{pqrs} = \{&\Pi_i\Pi_j\Pi_k\Pi_l \enskip | \enskip (i+j+k+l) \bmod 2 = 0, \label{eq:control-sector}\\
    &(i,j,k,l) \in \left(F(s,p), F(p,q), F(q,r), F(r,s)\right)\} \nonumber \ \ ,
\end{align}
where $0 \leq i,j,k,l \leq d-1$.
This form highlights the $D_4$ symmetry inherent to $\qdplaq$, which we leverage
to further optimize the classical preprocessing required to construct $\qdplaq$:
there are $u=\frac{c(c+1)(c^2+c+2)}{8}$ ways to assign $c=(d-1)$ choices of $\Gi$ to four links of a plaquette under $D_4$ symmetry, where $u$ is a doubly triangular number.
Thus, we only need to compute $\phi$ and $\mathcal{C}_{pqrs}$ for $u$ unique $\GiTerm$.
Then, the full set of $c^4$ $\GiTerm$ is obtained by permuting the indices $pqrs$.

Every pairing of operator $\mathcal{X}\mathcal{X}\mathcal{X}\mathcal{X}$ and $\PiTerm$ from the set of corresponding controls is associated with a non-zero physical transition matrix element of $\hat{\square}_\mathcal{P}$ (verified numerically for $d \leq 9$).  
As such, our construction provides a concise description of $\hat{\square}$ that
is applicable for any gauge field truncation, where $d \rightarrow \infty$ in the continuum limit.
For our demonstrated simulation, we specialize to a qutrit truncation ($d = 3, \Lambda_j = 1$).  
\Cref{tab:qutrit-plaq-op} enumerates the elements of the qutrit plaquette operator, denoted as $\hat{\square}^{(1)}$.  
To support future calculations with increasing truncation, the arbitrary-$d$
techniques developed above are applied to construct $\hat{\square}^{(3/2)}$ in Appendix~\ref{app:truncated-plaquette-operator}.

\begin{table}%
    \caption{\label{tab:qutrit-plaq-op}
        The $\PiTerm \GiTerm$ terms in the qutrit plaquette operator.
        Under $D_4$ symmetry, there are $u=6$ representative ways to assign $\Gi \in \{\Gi_{01}, \Gi_{12}\}$ to each $\GiTerm$.
        These representative classes are paired with $\PiTerm$ from the set $\mathcal{C}_{pqrs}$ defined in \Cref{eq:control-sector}.
        The control words $ijkl$ are shown, using the symbol {\es} to denote cases where all possible combinations of control values $0$ and $2$ are present.
        To obtain the complete set of $16$ $\GiTerm$, the $pqrs$ indices are permuted using $D_4$ transformations;  \enquote{order} 
        indicates the number of permutations resulting in a unique $pqrs$.
        The qutrit plaquette operator has 217 $\PiTerm \GiTerm$ terms, which are explicitly shown with $\phi$ values in \Cref{eq:full-qutrit-plaq-op}.
    }
    \begin{ruledtabular}
        \begin{tabular}{cc|cc}
        \multicolumn{2}{c}{$\Gi_{p p'} \Gi_{q q'} \Gi_{r r'} \Gi_{ss'}$} & \multicolumn{2}{c}{$\Pi_i\Pi_j\Pi_k\Pi_l$} \\
        $pqrs$ & order & $ijkl$ & $|\mathcal{C}_{pqrs}|$\\ 
        \hline
        $0000$ & 1 &
            \makecell[c]{
                $\begin{array}{cccc}
                    0000 & 0011 & 0101 & 0110 \\
                    1001 & 1010 & 1100 & 1111 
                \end{array}$
            } & 8  \\
        \hline
        $0001$ & 4 & 
            \makecell[c]{
                $\begin{array}{cccc}
                    1001 & 1012 & 1102 & 1111 \\
                    2002 & 2011 & 2101 & 2112 
                \end{array}$
            } & 8  \\
        \hline
        $0011$ & 4 & 
            \makecell[c]{
                $\begin{array}{cccc}
                    101\es & 1021 & 1111 & 112\es \\
                    2011 & 202\es & 211\es & 2121 
                \end{array}$
            } & 12  \\
        \hline
        $0101$ & 2 & 
            \makecell[c]{
                $\begin{array}{cccc}
                    1111 & 1122 & 1212 & 1221 \\
                    2112 & 2121 & 2211 & 2222 
                \end{array}$
            } & 8  \\
        \hline
        $0111$ & 4 & 
            \makecell[c]{
                $\begin{array}{cccc}
                    11\es\es  & 1111 & 12\es 1 & 121\es \\
                    21\es 1 & 211\es & 22\es\es & 2211
                \end{array}$
            } & 18  \\
        \hline
        $1111$ & 1 & 
            \makecell[c]{
                $\begin{array}{cccc}
                    \es\es\es\es & \es\es 11 & \es 1\es 1 & \es 11\es \\
                    1\es\es 1 & 1\es 1\es & 11\es\es  & 1111 
                \end{array}$
            } & 41  \\
        \end{tabular}
    \end{ruledtabular}
\end{table}

\subsection{Simulation Overview}
\label{subsec:sim-circuit}
\noindent
We apply a first-order Suzuki-Trotter decomposition with $N_T$ time steps to simulate real-time evolution of the gauge field on a cube.
\begin{figure*}[ht]
  \includegraphics[width=\textwidth]{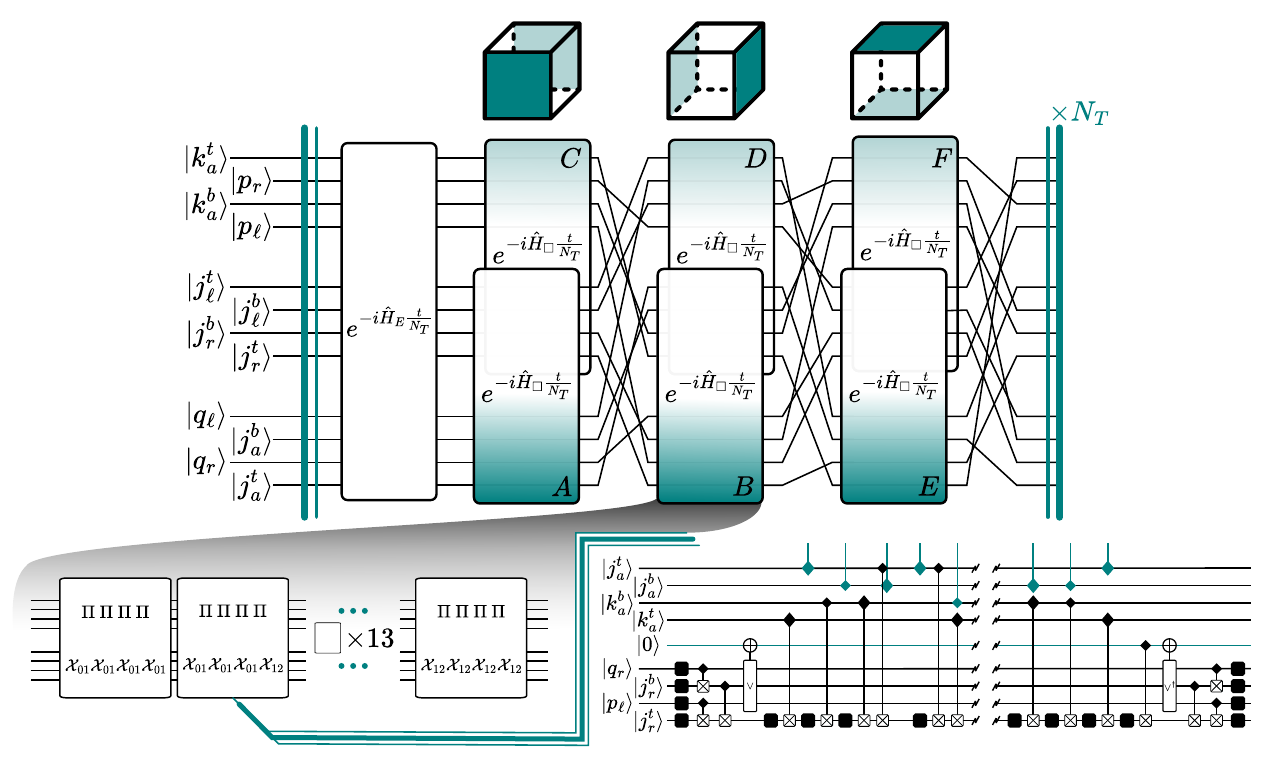}
  \caption{\label{fig:sim-circuit}
    Quantum circuit for a single Trotter step in the time evolution of SU(2) gauge fields on a cube.
    The simulation alternates between the Casimir operator and plaquette operators. 
    Evolution of the latter is parallelized in pairs of opposing faces.
    Each plaquette evolution is decomposed as a series of controlled $\GiTerm$ rotations.
    While those characterized by the same $\GiTerm$ rotation commute, an additional layer of Trotter error is introduced in ordering the $\GiTerm$ by increasing integer representation of their subspaces, i.e., the binary strings $pqrs$ in \Cref{tab:qutrit-plaq-op}.
    The structure of the elementary-gate level decomposition in the lower right corner is specific to the qutrit digitization of the gauge field.
    The boxed-\enquote{$\times$} circuit elements are qudit $X_{ij}$ operations (see \Cref{fig:gcx-gate}).
    }
\end{figure*}
A quantum circuit depicting the overarching structure of a Trotter step is presented in~\Cref{fig:sim-circuit}. 
Each step consists of evolution under the Casimir operator, followed by successive evolutions of the plaquette operator $e^{-i \hat{H}_\square \frac{t}{N_T}}$ over all six faces of the cube.

In the GVC of \Cref{eq:plaquetteGVC}, each plaquette operator evolution is decomposed into a series of controlled four-qudit rotations whose angles are proportional to physical transition matrix elements.
Our decompositions enable interleaving of controlled operations,
parallelizing $e^{-i \hat{H}_\square \frac{t}{N_T}}$ for opposite faces of the
cube that share control links, as depicted in the bottom right of \Cref{fig:sim-circuit}.
This approach leads to significant savings in circuit depth.
Similar parallelization opportunities arise for plaquette operators in genuine higher-dimensional lattices, whether organized by cubic components of a hypercubic lattice or residing in a broader class of spatial latticizations.

\section{Decomposition Techniques}
\label{sec:decompositions}
\noindent
The following decomposition techniques are presented in a general manner as they can be leveraged for qudit compilation beyond the present application which inspired their design.
We describe their application to efficiently synthesize the evolution of the plaquette operator for qutrit and qudit digitization of the gauge field in \Cref{subsec:qutrit-plaquette-decomposition} and \Cref{sec:qudit-decomp}, respectively.

We use the elementary qudit gate set of Ref.~\cite{Di:2013}, where 
\begin{figure} 
    \includegraphics[width=\columnwidth]{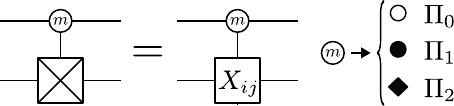}
    \caption{\label{fig:gcx-gate}The two-qudit GCX gate applies $X_{ij} = \ket{i}\bra{j} + \ket{j}\bra{i} + \sum_{k\notin \{i,j\}} \ket{k}\bra{k}$
    on the target qudit, conditioned on the control qudit being in state
    $\ket{m}$. For qutrits, we use distinct shapes to represent three possible control values.
    }
\end{figure}
the Generalized Controlled $X$ (GCX) gate (\Cref{fig:gcx-gate}) serves as our elementary two-qudit gate.
All GCX gates are assigned the same cost in our cost model, regardless of control value or $X_{ij}$ subspace.
In the present combination of SU(2) gauge field representation and decomposition techniques, nearly all GCX gates apply to nearest-neighbor qudit levels, which may be experimentally preferable.

\subsection{\UCDBT}
\label{subsec:ucdbt}
\noindent
The Hermitian operators $\mathcal{X}_{jk} = \ket{j}\bra{k} + \ket{k}\bra{j}$,
$\mathcal{Y}_{jk} = -i\ket{j}\bra{k} + i\ket{k}\bra{j}$, and $\mathcal{Z}_{jk} = \ket{j}\bra{j} - \ket{k}\bra{k}$ 
are the generators of two-level qudit rotation operators $R_\alpha(\theta) = \exp(-i\theta \alpha/2)$, $\alpha \in \{\mathcal{X},\mathcal{Y},\mathcal{Z}\}$ \cite{Di:2013}.
\begin{figure} 
    \includegraphics[width=\columnwidth]{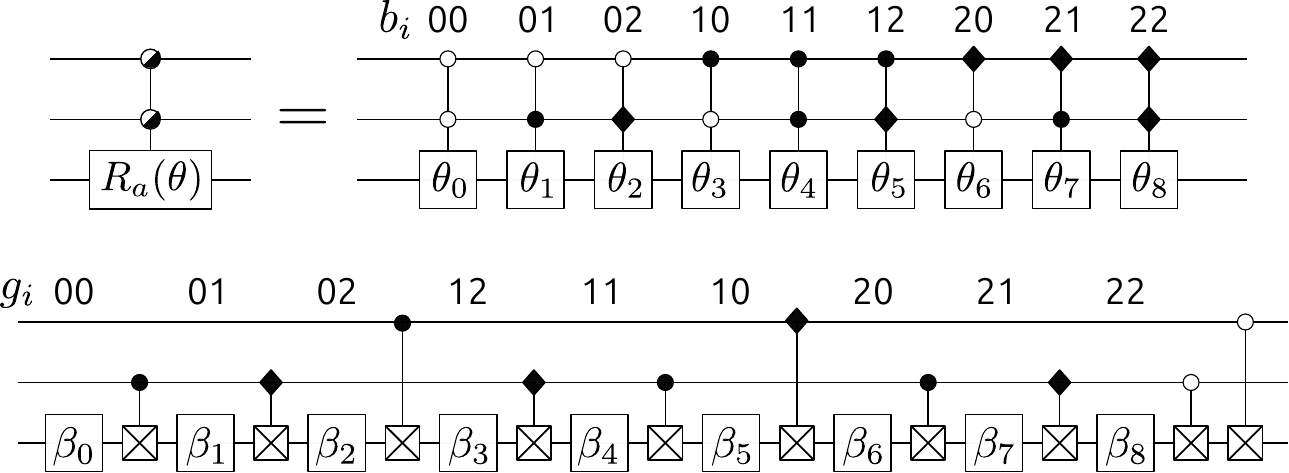}  
    \caption{\label{fig:qutrit-beta-trick}
    \textbf{(top)} {\UCDBT} $\ucnot{k}{d}{a}$ about the axis $a \in \{\mathcal{Y}, \mathcal{Z}\}$. 
    Above, the number of control qudits is $k=2$ and the qudit dimension is $d=3$.
    The gate label $\theta_i$ is shorthand for $R_{a}(\theta_i)$ (likewise for $\beta_i$).
    \textbf{(bottom)} The efficient decomposition of $\ucnot{2}{3}{a}$. 
    All gates are applied to the same subspace.
    }
\end{figure}
The {\UCDBT} $\ucnot{k}{d}{a}$ subroutine is the set of $k$-controlled $R_a$ gates, $a \in \{\mathcal{Y}, \mathcal{Z}\}$,
controlled on all $d^k$ possible control words\footnote{Consecutive $R_\Gi$ gates can be converted to consecutive
  $R_\mathcal{Z}$ gates using qudit Hadamard gates operating in the same subspace as
  the rotation gates. In all cases, the subspace of the rotation must be
  identical for all control words.
}.

We generalize existing techniques~\cite{Bergholm_2005,Mottonen:2005,Di:2015,Klco:2020} to efficiently decompose $\ucnot{k}{d}{a}$
by systematically constructing a circuit using the $d$-ary reflected Gray code.
\Cref{fig:qutrit-beta-trick} shows our technique applied to $\ucnot{2}{3}{a}$.
The index of the varying digit in the Gray code identifies the GCX control qudit; 
the control value is determined by the higher value of the varied digit.
As an example, the transition $12\rightarrow 11$ in the Gray code places a GCX
gate controlled on the second control qudit with control value 2.

Referring to the notation in \Cref{fig:qutrit-beta-trick}, the rotation angles $\bm{\theta}$ are transformed into $\bm{\beta}$ according to
\begin{align}
    \begin{bmatrix}
        \theta_0 \\ \vdots \\ \theta_{d^k-1}
    \end{bmatrix}
    = M \begin{bmatrix}
        \beta_0 \\ \vdots \\ \beta_{d^k-1}
    \end{bmatrix}, &&
    M_{ij} = \prod_{l=0}^{k-1} (-1)^{f(b_l^{(i)}, g_l^{(j)})}, \nonumber \\
    f(b_l^{(i)},g_l^{(j)}) &&= \begin{cases} 1, & 0 < b_l^{(i)} \leq
                                                  g_l^{(j)}\\ 0,
                                                &\hbox{otherwise} \end{cases}, \label{eq:qudit-beta-matrix}
\end{align}
where $b_l^{(m)}$ is the $l$-th digit of the $d$-ary representation of the integer $m$, and $g_l^{(m)}$ is the $l$-th digit of the $d$-ary reflected Gray code term at index $m$.
We remark that for qubits, $f$ evaluates to 1 only when both inputs are 1, and the sum over all bit indices reduces to a simple dot product of the binary and Gray code sequences for $i$ and $j$. 
This yields exactly the inverse of the expression for $M_{ij}$ in Equation 3 of Ref.~\cite{Mottonen:2005}. 
For $d>2$, \Cref{eq:qudit-beta-matrix} is no longer simply a dot product as it depends on the order of the arguments.

This decomposition of $\ucnot{k}{d}{a}$ requires $d^k$ single-qudit rotation gates, at most $(d^k + k - 1)$ GCX gates, and at most one $X_{ij}$ gate.
For a full account of the circuit construction and resource count, as well as a ququart example, see Appendix~\ref{app:ucdbt}.

To the best of our knowledge, the most efficient known decomposition for  $\ucnot{k}{d}{a}$ requires $2d^{k-1}(d-1)$ GCX gates \cite{Di:2013}.
In the present application where $k$ is fixed by the lattice connectivity and
$d$ increases with gauge field truncation, our decomposition is a $2 \times$
improvement on the GCX gate count and depth in the limit that $d \rightarrow \infty$.

\subsection{\FredkinTrick}
\label{subsec:fredkintrick}
\noindent
Let $d_1$ and $d_2$ be two qudit dimensions, $d_2 > d_1$.
We implement a technique, {\FredkinTrick}, that allows the use of $d_1$-dimensional qudit decompositions as subroutines in $d_2$-dimensional qudit circuits, provided a ``gating subcircuit" is constructed.
The gating subcircuit prevents a $d_1$-dimensional subroutine from being applied when inputs to the subroutine are outside the $d_1$-dimensional subspace.
Essentially, {\FredkinTrick} uses an auxiliary qudit to exclude spectator dimensions, leading to reduced gate counts.

\begin{figure}
    \includegraphics[width=\columnwidth]{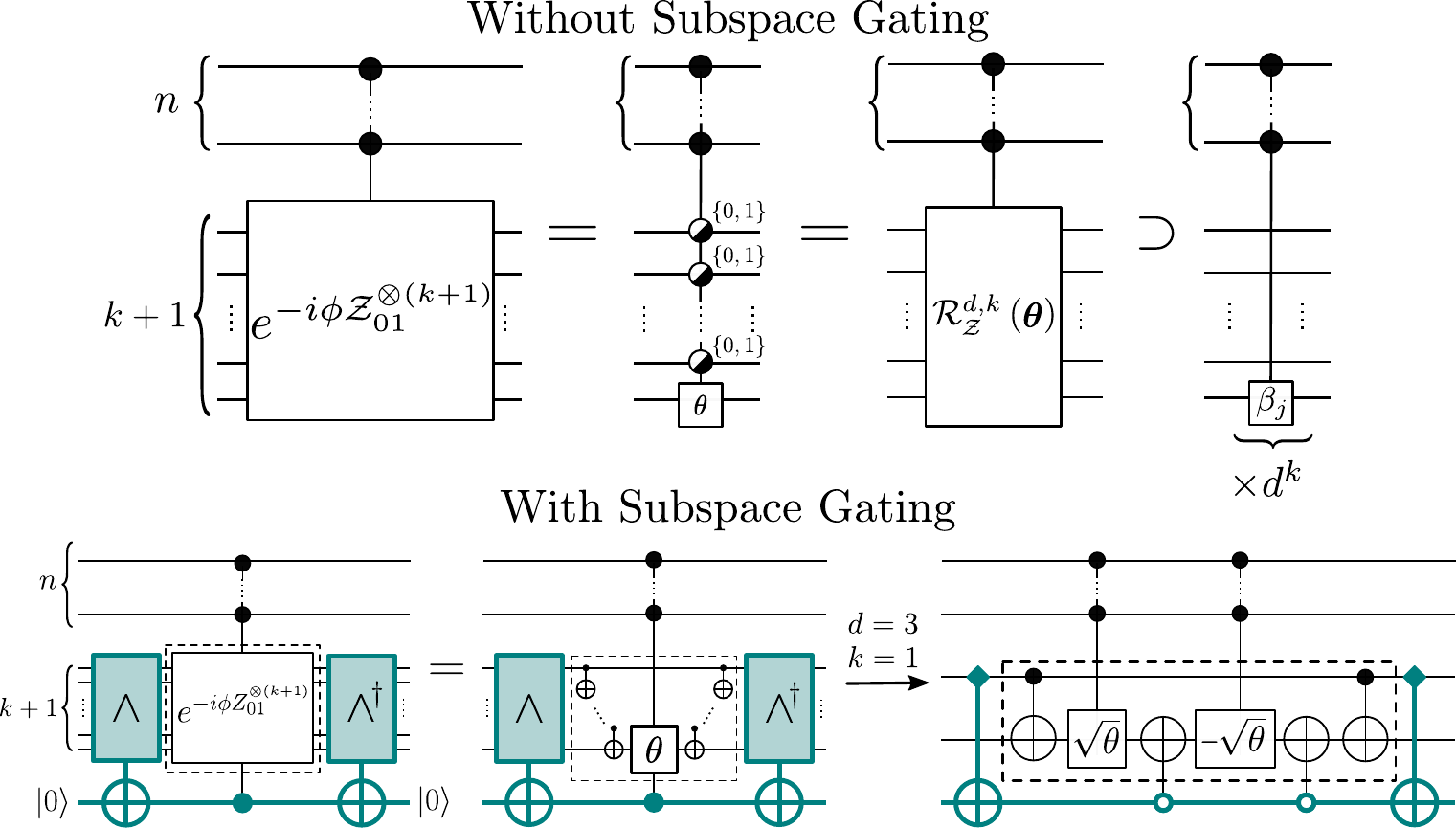}
    \caption{\label{fig:or_fredkin_trick}
    The unitary $\exp(-i\phi\Pi^{\otimes n}\mathcal{Z}^{\otimes k+1}_{01})$ 
    applies $\exp(-i\phi \mathcal{Z}^{\otimes k+1}_{01})$ controlled by an $n$-qudit register.
    \textbf{(top)} Without {\FredkinTrick}, $\exp(-i\phi \mathcal{Z}_{01}^{\otimes k+1})$
    is decomposed as a sequence of $k$-controlled $R_{\mathcal{Z}_{01}}$ gates with control words spanning $\mathbb{Z}_2^k$.
    This is equivalent to $\ucnot{k}{d}{{\mathcal{Z}_{01}}}$ where $\theta \in \bm{\theta}$ associated with control words in $\mathbb{Z}_d^k \setminus \mathbb{Z}_2^k$ are $0$. 
    Decomposing $\ucnot{k}{d}{{\mathcal{Z}_{01}}}$ with the technique in \Cref{subsec:ucdbt} yields $d^k$ rotation gates,
    each controlled by the upper register, yielding a total of $d^k$ $n$-controlled rotation gates.
    \textbf{(bottom)} With {\FredkinTrick}, the gating subcircuit (teal) enables
    the decomposition of $\exp(-i\phi \mathcal{Z}^{\otimes k+1}_{01})$ in the
    qubit subspace, i.e., the decomposition of $\exp(-i\phi Z^{\otimes
      k+1}_{01})$. This requires two $n$-controlled rotation gates regardless of $k$.
    On the right, we show the small case of $d=3$, and $k=1$ for concreteness. 
    }
\end{figure}

\Cref{fig:or_fredkin_trick} shows an instance where {\FredkinTrick} saves GCX gates.
For comparison, we will temporarily use the number of $n$-controlled rotation gates as a proxy for the number of GCX gates.
The unitary $\exp(-i\phi\Pi^{\otimes n}\mathcal{Z}^{\otimes k+1}_{01})$ can be expressed as an $n$-controlled $\ucnot{k}{d}{\mathcal{Z}}$, yielding $d^k$ $n$-controlled rotation gates via the decomposition in \Cref{subsec:ucdbt}.
By contrast, with {\FredkinTrick}, $\exp(-i\phi\Pi^{\otimes n}\mathcal{Z}^{\otimes k+1}_{01})$ decomposes into two $n$-controlled rotation gates, independent of both $d$ and $k$.
The savings stem from use of the qubit subroutine $\exp(-i\phi Z^{\otimes k+1}_{01})$ (bottom of \Cref{fig:or_fredkin_trick}) enabled by the gating subcircuit.

Generally speaking, the gating subcircuit verifies an input is in the subroutine's subspace using the $\land$ (AND) gate before toggling the auxiliary qudit to activate the subroutine.
In the special case of qutrits, it is more efficient to implement the $\lor$ (OR) gate version of the input verification gate (which is obtained using De Morgan's law \cite{de1847formal}, described in \Cref{fig:de-morgan}).
An example is given in the qutrit specialization presented at the bottom right of \Cref{fig:or_fredkin_trick}.
In this case, the $\lor$ gate is implemented as a single control-on-$\ket{2}$ GCX gate.
No gates are required for the bottom input qutrit since the subroutine naturally has no effect when the bottom qutrit is in state $\ket{2}$, a subroutine-specific optimization we apply throughout this work.
We provide input verification gates for arbitrary $d$ and $k$ in \Cref{app:fredkin-trick}.

As a caveat, we note there exists a better decomposition of
$\exp(-i\phi\Pi^{\otimes n}\mathcal{Z}^{\otimes k+1}_{01})$ that takes advantage
of our specific $\theta$ values (see \Cref{fig:alt-angle-decomp}), yielding $2^k$ $n$-controlled rotation gates.
Although $2^k$ improves upon $d^k$, it is still worse than the 2 achievable through {\FredkinTrick}.

\subsection{\CCDBT}
\label{subsec:ccdbt}
\noindent
Let $\ctrlseq$ be a sequence of control words ordered by increasing integer representation of their $d$-ary strings.
For $\ucnot{k}{d}{a}$ described in \Cref{subsec:ucdbt}, the control sequence is $\{\ctrlseq\}=\mathbb{Z}_d^k$.
We generalize $\ucnot{k}{d}{a} \rightarrow \ccnot{k}{d}{a}$ where $\ctrlseq
\subseteq \mathbb{Z}_d^k$, i.e., $\ctrlseq$ does not contain all possible control words,
under the condition that the control qudits are also restricted to $\ket{i \in \{\ctrlseq\}}$.
Qudit states may be restricted in circuits with mixed-dimensional qudits, artificially using {\FredkinTrick}, or due to the properties of the quantum simulation.
\begin{figure}
    \includegraphics[width=\columnwidth]{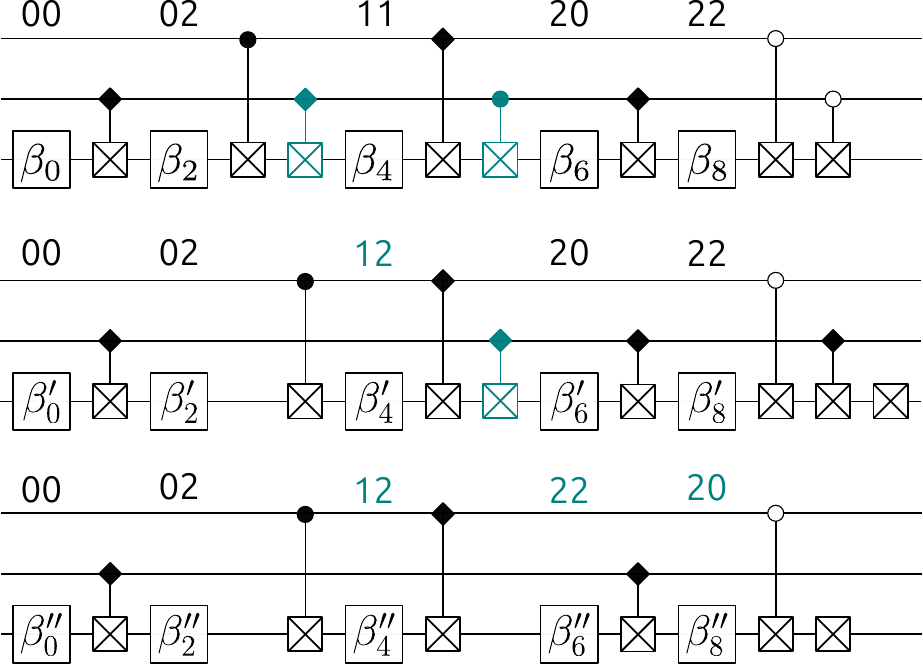}
    \caption{\label{fig:ccdbt-example}
    A step-by-step decomposition of $\ccnot{2}{3}{\mathcal{Z}}$ with $\ctrlseq=(00,02,11,20,22)$.
    We begin with the Gray-code ordered $\ctrlseq$ (which in this particular
    case is equivalent to $\ctrlseq$) and generate the quantum circuit using the procedure described in \Cref{subsec:ucdbt}.
    \textbf{(top)} 
    Assuming the control qutrits are restricted to $\ket{i\in \{\ctrlseq\}}$, observe that the top control qutrit being in state
    $\ket{1}$ uniquely identifies the control word ``11'' and its corresponding
    rotation angle $\theta_4$.
    The gates coloured in teal are not necessary, and we are free to relabel the Gray code term from ``11'' to ``12''. Doing so saves one GCX gate.
    The relabelling changes the $\beta$s to $\beta^\prime$s, while the change is also reflected in $M$ such that the overall operation is unchanged. 
    \textbf{(middle)} Since $\theta$s are sums of $\beta$s, the order in which their constituent rotations occurs is arbitrary.
    Thus, we are free to swap the order of ``20'' and ``22'', as this saves another GCX gate.
    \textbf{(bottom)} The final decomposition of $\ccnot{2}{3}{\mathcal{Z}}$ applies only one GCX gate between rotation gates.
    }
\end{figure}
\Cref{fig:ccdbt-example} shows an example of $\ccnot{2}{3}{\mathcal{Z}}$ where $\ctrlseq$ contains the even-numbered control words spanning two control qutrits.

We provide a decomposition of $\ccnot{k}{d}{a}$ in Appendix~\ref{app:ccdbt} for arbitrary $d$, $k$, and $\ctrlseq$.
Our implementation generates two sequences of $d$-ary strings, $b^{\prime}$ and $g^{\prime}$, from $\ctrlseq$.
The sequences $b^{\prime}$ and $g^{\prime}$ are used to transform the angles $\bm{\theta} \rightarrow \bm{\beta}$ according to \Cref{eq:qudit-beta-matrix}
(taking the shorter sequence length into account when constructing $M$) and $g^\prime$ generates the quantum circuit.

By keeping $\ctrlseq$ general, we sacrifice the guarantee that the matrix $M$ is invertible,
although we conjecture our particular implementation yields an invertible $M$ when
\begin{equation}
    \{\ctrlseq\} = \mathbb{Z}_{d_1} \times \mathbb{Z}_{d_2} \times \cdots \times \mathbb{Z}_{d_k},
    \label{eq:conjecture}
\end{equation}
i.e., uniform control but non-uniform qudit dimension.

Although our implementation produces the decomposition in \Cref{fig:ccdbt-example} with an invertible $M$,
we encountered cases where this did not hold, specifically for larger $\bm{C}$ comprising all even-numbered control words for a mixed-dimensional qudit system.
An example is found in \Cref{sec:qutrit-cube-sim} (\Cref{eq:control-sequence}).
In that situation, manual corrections were applied (as described in \Cref{app:corr-82-ccdbt}) to make $M$ invertible.
Although the manual corrections may not be optimal, their existence suggests that there is an improved implementation which does yield an invertible $M$ under such control sequence restrictions.
Finding this implementation, which we henceforth refer to as the ``hypothesized decomposition'' is the subject of future work.

\section{Qutrit Cube Simulation}
\label{sec:qutrit-cube-sim}
\noindent
We apply the decomposition techniques described in \Cref{sec:decompositions} to simulate SU(2) gauge field on a cube, truncated to $\Lambda_j = 1$.
First, we decompose the Trotterized unitary evolution of the plaquette operator digitized to qutrits.
Then, we repeatedly apply the decomposed operator to perform the qutrit cube simulation.

\subsection{Plaquette Operator Evolution}
\label{subsec:qutrit-plaquette-decomposition}
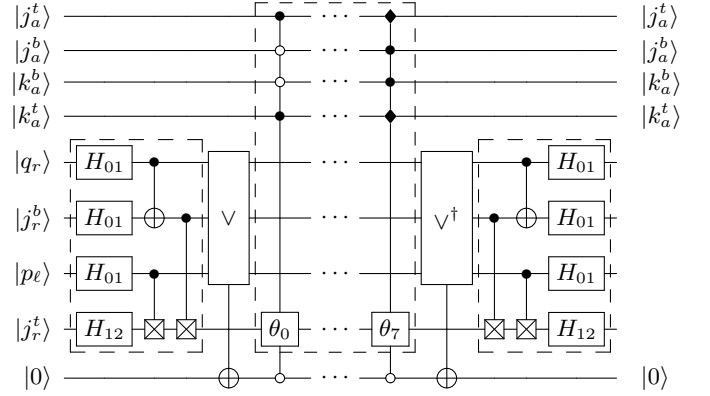
\begin{figure}
    \[
        \Qcircuit @C=.5em @R=.9em {
            & \lstick{|j^t_a\rangle} & \qw & \qw & \qw & \qw & \tctrlo{1} \gategroup{1}{7}{8}{13}{.5em}{--} & \qw &  & \cdots &  &  & \tctrlt{1} & \qw & \qw & \qw & \qw & \qw & \rstick{|j^t_a\rangle} \\ 
            & \lstick{|j^b_a\rangle} & \qw & \qw & \qw & \qw & \tctrlz{1} & \qw &  & \cdots &  &  & \tctrlo{1} & \qw & \qw & \qw & \qw & \qw & \rstick{|j^b_a\rangle} \\ 
            & \lstick{|k^b_a\rangle} & \qw & \qw & \qw & \qw & \tctrlz{1} & \qw &  & \cdots &  &  & \tctrlo{1} & \qw & \qw & \qw & \qw & \qw & \rstick{|k^b_a\rangle} \\ 
            & \lstick{|k^t_a\rangle} & \qw & \qw & \qw & \qw & \tctrlo{4} & \qw &  & \cdots &  &  & \tctrlt{4} & \qw & \qw & \qw & \qw & \qw & \rstick{|k^t_a\rangle} \\ 
            & \lstick{|q_r\rangle} & \gate{H_{01}} \gategroup{5}{3}{8}{5}{.5em}{--} & \ctrl{1} & \qw & \multigate{2}{\lor} & \qw & \qw &  & \cdots &  &  & \qw & \multigate{2}{\lor^\dagger} & \qw \gategroup{5}{15}{8}{17}{.5em}{--} & \ctrl{1} & \gate{H_{01}} & \qw &  \\ 
            & \lstick{|j^b_r\rangle} & \gate{H_{01}} & \targ & \ctrl{2} & \ghost{\lor} & \qw & \qw &  & \cdots &  &  & \qw & \ghost{\lor^\dagger} & \ctrl{2} & \targ & \gate{H_{01}} & \qw &  \\ 
            & \lstick{|p_\ell\rangle} & \gate{H_{01}} & \ctrl{1} & \qw & \ghost{\lor} & \qw & \qw &  & \cdots &  &  & \qw & \ghost{\lor^\dagger} & \qw & \ctrl{1} & \gate{H_{01}} & \qw &  \\ 
            & \lstick{|j^t_r\rangle} & \gate{H_{12}} & \boxtarg & \boxtarg & \qw \qwx & \gate{\theta_0} & \qw &  & \cdots &  &  & \gate{\theta_7} & \qw \qwx & \boxtarg & \boxtarg & \gate{H_{12}} & \qw &  \\ 
            & \lstick{|0\rangle} & \qw & \qw & \qw & \targ \qwx & \ctrlo{-1} & \qw &  & \cdots &  &  & \ctrlo{-1} & \targ \qwx & \qw & \qw & \qw & \qw & \rstick{|0\rangle}
        }
    \]
    \caption{\label{fig:qutrit-circuit}
        Decomposition of $ \exp(-i \tau \sumpi
        \Gi_{01}\Gi_{01}\Gi_{01}\Gi_{12})$ applied to plaquette B of the cube
        (equivalent to the bottom-right of \Cref{fig:sim-circuit}, without parallelization).
        The boxed-``$\times$'' is $X_{12}$ whereas the regular CNOT target represents $X_{01}$.
        The circuit evolves the plaquette links while leaving the control links and auxiliary qutrit unchanged.
        The subroutine on the left calculates the subspace parity in the computational basis, so we name it the $\mathcal{X}$-parity subroutine.
        The subroutine in the middle is a sequence of four-qutrit controlled $R_{\mathcal{Z}_{12}} (\theta_i)$  with control words corresponding to row 2 of \Cref{tab:qutrit-plaq-op}.
        An additional control is placed on the rotations by the gating subcircuit, which gates the subroutine 
        by setting the auxiliary qutrit to $\ket{1}$ if any of the first three input qutrits are in state $\ket{2}$.
        The subroutine naturally has no effect if the fourth input qutrit $\ket{j^t_r}$ is in state $\ket{0}$, as all gates operating on that qutrit are in the $(1, 2)$ subspace.
        Therefore, only the three-input $\lor$ gate is needed to subspace gate the plaquette links.
        }
\end{figure}

\noindent
The Trotterized evolution of the qutrit plaquette operator is
\begin{equation}
    e^{-i\tau \qtplaq} \stackrel{\text{Trot}}{=} \prod_{\GiTerm} e^{-i \tau \sumpi \GiTerm},
    \label{eq:trotterized-qutrit-plaq}
\end{equation}
where $\sumpi = \left(\sum_{c \in \mathcal{C}_{pqrs}} \phi c\right)$ is shorthand for a $\phi$-weighted sum over all projectors 
associated with the term $\Gi_{pp^\prime}\Gi_{qq^\prime}\Gi_{rr^\prime}\Gi_{ss^\prime}$ (\Cref{eq:control-sector} and $ijkl$ column in \Cref{tab:qutrit-plaq-op})
and $\tau = \minus \frac{t}{g^2N_T}$ incorporates constants from \Cref{eq:yang-mills}.
We take the following steps to decompose $\uqtplaq$:
\begin{enumerate}
    \item decompose $\onexxxxunitary$ into the circuit in \Cref{fig:qutrit-circuit} with {\FredkinTrick} (\Cref{subsec:fredkintrick}),
    \item decompose the remaining {\DBT} into elementary qutrit gates (\Cref{subsec:ccdbt}),
    \item apply the decomposed circuit $16$ times, varying parameters for each $\GiTerm$, to implement $\uqtplaq$.
\end{enumerate}

In step 1, we decompose $\onexxxxunitary$ by {\FredkinTrick} the plaquette links to enable the qubit decomposition of $\exp(-i \tau \sumpi XXXX)$.
We factor out Hadamard gates ($H$) and GCX gates to obtain the circuit in \Cref{fig:qutrit-circuit}.
The subspaces of $H_{ij}$ and $X_{ij}$ are the same as those of $\Gi_{ij}$
assigned to the plaquette links, and the control values of the GCX gates are the
larger of $i$ and $j$.
The gating subcircuit uses the three-input $\lor$ gate (\Cref{fig:three-input-lor}) to verify that the plaquette links are in the same subspace as $\Gi_{pp^\prime}\Gi_{qq^\prime}\Gi_{rr^\prime}\Gi_{ss^\prime}$.
Lastly, we commute the $\lor$ gates through the $H$ and GCX gates to facilitate further cancellations.
At this stage, all gates are elementary qutrit gates except for the contiguous $(n+1)$-controlled rotations in the middle of the circuit.

In step 2, we decompose the {\CCDBT} using the technique in \Cref{subsec:ccdbt}.
These rotations are controlled by five qutrits (four control links and one auxiliary qutrit from the gating subcircuit).
The control words corresponding to the control links are constrained by $(i+j+k+l)\bmod 2=0$ from \Cref{eq:control-sector},
and the control value of the auxiliary qutrit is restricted to $\{0, 1\}$.
Thus, the control sequence of $\ccnot{5}{3}{\mathcal{Z}}$ is 
\begin{equation}
    \{\ctrlseq\} = \{w \in \mathbb{Z}_2 \times \{v \in \mathbb{Z}_3^4 \enskip \mid v \bmod 2 = 0\}\},\ |\ctrlseq|=82.
    \label{eq:control-sequence}
\end{equation}
In this case, the matrix $M$ is not invertible and requires manual correction (via additional GCX gates).
The corrected version, $\manccnot{5}{3}{\mathcal{Z}}$, incurs $12$ additional GCX gates and $12$ additional depth (see \Cref{app:corr-82-ccdbt}).
Despite this, $\manccnot{5}{3}{\mathcal{Z}}$ uses $94$ GCX gates, far less than
the  $(3^5-1)+5=247$ used for $\ucnot{5}{3}{\mathcal{Z}}$.

Finally, in step 3, we implement $\uqtplaq$ as per
\Cref{eq:trotterized-qutrit-plaq} by applying circuits like that in \Cref{fig:qutrit-circuit} for each of the $16$ distinct $\GiTerm$.
The subspaces, control values of the qutrit gates, and the $\bm{\theta}$ angles
are changed in accordance with \Cref{tab:qutrit-plaq-op}.
\begin{table}
    \caption{\label{tab:gggg-decomp-resource}
        Quantum resource estimates for the Trotterized evolution of one qutrit plaquette operator $\qtplaq$.
        The Trotterized $\uqtplaq$ is a product of $16$ unitaries of the form $\onexxxxunitary$, denoted as $\sumpi \GiTerm$ in the second last row.
        The decomposition uses one auxiliary qutrit, leading the operator to span a total of nine qutrits.
    }
    \begin{ruledtabular}
        {\def\arraystretch{1.3}
        \begin{tabular}{c|c|cccc|c}
        Subcircuits      & Counts & GCX & $R_\mathcal{Z}$ & $X$ & $H$ & Depth \\
        \hline 
        $\manccnot{5}{3}{\mathcal{Z}}$ & $\times$1 & 94  & 82  & 0  & 0 & 176  \\
        $\lor$ gate & $\times$2  & 6   & 0   & 1  & 0 & 6    \\
        $\mathcal{X}$-parity & $\times$2  & 3   & 0   & 0  & 4 & 3    \\
        \hline
        $\left(\sum_\Pi\right) \GiTerm$        &  $\times$16   & 112  & 82   & 2   & 8  & 194\\ \hline
        $\qtplaq$ (Total)       &  & 1792 & 1312 & 32  & 128& 3104
        \end{tabular}}
    \end{ruledtabular}
\end{table}
The quantum resource estimates for implementing one $\uqtplaq$ are shown in \Cref{tab:gggg-decomp-resource}.

As a final remark, we developed an alternate decomposition of $\uqtplaq$ that
leverages the $(i+j+k+l)\bmod 2=0$ constraint on control words, in
addition to \Cref{eq:control-set}.
For clarity of presentation, we relegate this more detailed decomposition to Appendix~\ref{app:gggg-decomp-alt}.

\subsection{Cube Simulation}
\label{sec:resource-scaling}
\begin{figure}
    \includegraphics[width=\columnwidth]{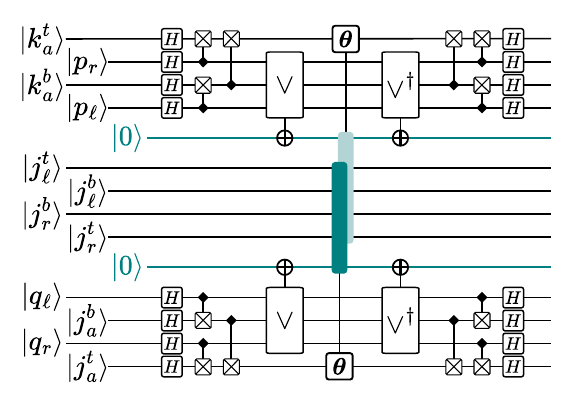}
    \caption{\label{fig:gggg-decomp}
    Decomposition of two $\onexxxxunitary$, applied in parallel to opposite faces of the cube, using two auxiliary qutrits to apply {\FredkinTrick} on the plaquette links.
    The subspaces and control values of all gates are contingent on the particular $\GiTerm$ term.
    }
\end{figure}

\noindent
To simulate the real-time dynamics of SU(2) gauge fields on a cube, we repeatedly apply the decomposed $\uqtplaq$ as shown in \Cref{fig:sim-circuit}.
Taking advantage of the parallelization opportunity described in
\Cref{subsec:sim-circuit}, we stack two $\uqtplaq$ decompositions applied to
opposite faces of the cube to halve the depth of the circuit, as shown in \Cref{fig:gggg-decomp}. 
We reverse the gate order for one face and interleave the controls on the shared gauge links (bottom-right of \Cref{fig:sim-circuit}).
Simultaneously evolving two faces requires a second auxiliary qutrit,
increasing the qutrit count of the cube simulation with {\FredkinTrick} to
14.

\begin{figure}
    \includegraphics[width=\columnwidth]{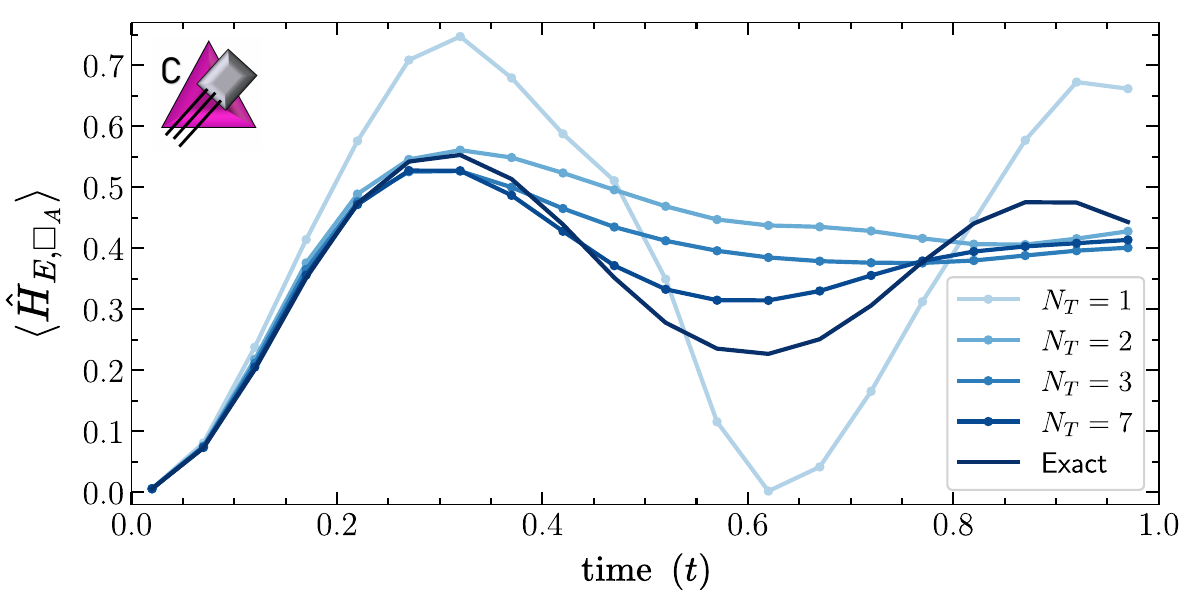}
    \caption{\label{fig:results}
        Expectation value of the electric energy contribution of plaquette A on the cube with coupling $g^2=0.2$.
        The Trotterized results converge towards the exact time evolution, which was simulated in the global basis of physical states (see \Cref{app:simulation-results}).
        The icon on the top left, introduced in Ref~\cite{Klco:2020}, denotes purely classical simulation.
    }
\end{figure}

Resources for one Trotter step of the cube simulation are shown in \Cref{tab:cube-sim-resource}.
We implemented the simulation in PennyLane \cite{pennylane}, verifying the correctness of our decompositions by comparing simulation observables (\Cref{fig:results})
with exact numerical results.
The simulation was executed on a workstation running Ubuntu 24.04 with an Intel Core i7-12700KF 3.6 GHz CPU, 96GB RAM, and MSI GeForce RTX 4090 GPU with 24GB memory. 
A custom PennyLane (branched from version 0.38)~\cite{custom-pennylane} with specialized qutrit operations was used, 
alongside GPU-enabled JAX (v-0.4.12) and just-in-time compilation. 
It took roughly 3 hours to generate data for \Cref{fig:results}.

PennyLane was also used to obtain quantum resource estimates directly from the implemented circuits.
We found exact agreement on all values in \Cref{tab:cube-sim-resource} except
depth, which was reduced from 9314 to 9284.
We attribute this improvement to PennyLane parallelizing additional gates, e.g., at the boundaries between distinct $\GiTerm$.
Compiler optimizations, such as cancelling adjacent self-inverse gates, merging adjacent rotation gates,
and Hadamard-GCX rewrite rules (Appendix~\ref{app:hadamard-gcx}) can be used to further reduce gate count and circuit depth.

\begin{table}
  \caption{\label{tab:cube-sim-resource} 
    Resource estimates for one full Trotter step of the cube simulation (\Cref{fig:sim-circuit}).
    Evolution of two $\qtplaq$s (row 1) on opposite cube faces occurs in parallel.
    As such, the gate counts are double that of a single $\qtplaq$ (last row of \Cref{tab:gggg-decomp-resource}) while the depth remains the same.
    Repeating the parallelized face evolution thrice to evolve all six faces,
    and incorporating the electric operator $\hat{E}^2$ (row 2) gives the total resource estimates for one Trotter step (last row).
    The decomposition requires two auxiliary qutrits, and 14 qutrits in total.
    }
    \begin{ruledtabular}
        {\def\arraystretch{1.2}
        \begin{tabular}{c|c|cccc|c}
          Operators & Counts & GCX & $R_\mathcal{Z}$ & $X$ & $H$ & Depth \\
          \hline
            two $\qtplaq$ & $\times$3 & 3584  & 2624 & 64  & 256 & 3104 \\
            $\hat{E}^2$   & $\times$1 &     0 &   24 &   0 &   0 &    2 \\
            \hline
            Total         &    & 10752 & 7896 & 192 & 768 & 9314 \\
        \end{tabular}}
    \end{ruledtabular}
\end{table}

\section{Qudit Simulations and Resource Scaling}
\label{sec:qudit-decomp}
\noindent
To simulate SU(2) non-Abelian gauge fields at arbitrary truncation level $\Lambda_j = (d-1)/2$,  we digitize to qudits, $d > 3$.
The procedure for decomposing Trotterized unitary evolution of $\uqdeplaq$ is similar to the qutrit case in \Cref{subsec:qutrit-plaquette-decomposition}, with key differences detailed below.

In step 1, we use the $\land$-version of the input verification gate, shown in \Cref{fig:input-verification}, instead of the $\lor$-version used for the qutrit case.
Since the $\land$ gate contains gates operating in the same subspace as the qubit subroutine, we can no longer commute the $\land$ gate through the $\Gi$-parity subroutine as we did for the $\lor$ gate in \Cref{fig:qutrit-circuit}.
Despite missing out on some cancellations, the $\land$ gate is more efficient overall than the $\lor$ gate when $d>3$.
We further observe that the dimension of the auxiliary qudit is independent of gauge-field truncation level, scaling instead with the lattice connectivity.
With the current connectivity, the auxiliary is a ququart, reflecting the {\FredkinTrick} of four plaquette links.

In step 2, we decompose the {\CCDBT} $\ccnot{5}{d}{\mathcal{Z}}$ with the control sequence
\begin{equation}
    \{\ctrlseq\} = \{w \in \mathbb{Z}_2 \times \mathbb{Z}_3^4\}
\end{equation}
instead of \Cref{eq:control-sequence}.
Doing so forgoes the resource improvement from the naturally restricted \Cref{eq:control-sequence}, effectively doubling the number of GCX and $R_\mathcal{Z}$ gates,
but with the benefit of an invertible $M$.

In step 3, we repeatedly apply the decomposed circuit $(d-1)^4$ times (varying parameters appropriately), to implement $\uqdeplaq$.
\begin{table*} 
    \caption{\label{tab:qudit-decomp-resource}
    Upper bound resource estimates for time evolution of one plaquette operator $\qdplaq^{\Lambda_j=(d-1)/2}$ (last two rows).
    The organization of the table follows that of \Cref{tab:gggg-decomp-resource}.
    We use the decomposition of $\ccnot{5}{d}{\mathcal{Z}}$ with $\{\ctrlseq\} = \mathbb{Z}_2 \times \mathbb{Z}_d^4$ (to avoid the need for manual corrections)
    and the $\land$ gate (\Cref{fig:input-verification}) in our gating subcircuit.
    A single auxiliary qudit is used in the decomposition, leading the operator to span a total of nine qudits.
    }
    \begin{ruledtabular}
        {\def\arraystretch{1.4}
        \begin{tabular}{c|c|cccc|c}
        Subcircuits                     & Counts                & GCX         & $R_\mathcal{Z}$     & $X$ & $H$ & Depth \\ 
        \hline 
        \hline 
        $\ccnot{5}{d}{\mathcal{Z}}$     & $\times$1             & $2d^4+4$    & $2d^4$    & 1  & 0 & $4d^4+5$  \\
        $\land$ gate                    & $\times$2             & 10          & 0         & 0  & 0 & 10   \\
        $\mathcal{X}$-parity            & $\times$2             & 3           & 0         & 0  & 4 & 3    \\
        \hline
        $\sumpi \GiTerm$                       & $\times(d-1)^4$       & $2d^4+30$   & $2d^4$     & 1  & 8 & $4d^4+31$\\
        $\qdplaq$ (Total)               &                       & $(d-1)^4(2d^4+30)$    & $2(d-1)^4d^4$  & $(d-1)^4$ & $8(d-1)^4$ & $(d-1)^4(4d^4+31)$\\
        \hline
        $\qdplaq$ (Scaling)             &                       & $O(d^8)$    & $O(d^8)$  & $O(d^4)$ & $O(d^4)$ & $O(d^8)$
        \end{tabular}}
    \end{ruledtabular}
\end{table*}
We report resource estimates for the full $\uqdeplaq$ in \Cref{tab:qudit-decomp-resource}.
We achieved an overall GCX gate count of $(d-1)^4(2d^4+30)$, and circuit depth $(d-1)^4(4d^4+31)$.
Substituting $d=3$ and comparing with the qutrit resources in \Cref{tab:gggg-decomp-resource}, we find only expected discrepancies due to the different gating subcircuits and the $\ccnot{5}{d}{\mathcal{Z}}$ decomposition.

The resource estimates can be related directly to the plaquette operator GVC.
In the following description, we consider access to the hypothesized decomposition of $\ccnot{k}{d}{a}$ discussed in \Cref{subsec:ccdbt}.
With the aid of one auxiliary ququart, the presented decomposition uses GCX gates and $R_\mathcal{Z}$ gates with counts roughly equal to the number of $\PiTerm \GiTerm$ combinations, constrained
only by the qudit dimension $d$ and not by \Cref{eq:control-sector}.
This connection originates from the matrix $M$ transforming $\bm{\theta}$ of size $|\mathcal{C}_{pqrs}|$ to $\bm{\beta}$ of size $|\ctrlseq|$ for each $\Gi_{pp^\prime}\Gi_{qq^\prime}\Gi_{rr^\prime}\Gi_{ss^\prime}$,
where $\ctrlseq$ corresponds to the qudit version of \Cref{eq:control-sequence}.
These counts are half of those reported in \Cref{tab:qudit-decomp-resource}, bringing the qudit resource estimates inline with those for qutrits (\Cref{tab:gggg-decomp-resource}).
\Cref{tab:qudit-decomp-resource} serves as a resource upper bound in the absence of the hypothesized decomposition of $\ccnot{k}{d}{a}$.
The hypothesized decomposition precludes manual corrections in the qutrit case as well.

The alternate decomposition presented in \Cref{app:gggg-decomp-alt} achieves better scaling in gate count and depth at the cost of three additional auxiliary qudits.
However, if we free the dimension of the auxiliary from the gauge-field truncation as we did above, then the decomposition can be done with one qu8it, an 8-dimensional qudit,
reflecting the {\FredkinTrick} of all eight gauge field links.
The alternate decomposition uses GCX gates and $R_\mathcal{Z}$ gates with counts roughly double the number of non-zero transition matrix elements, as obtained by \Cref{eq:control-sector}.
This scaling can be seen in the resource estimates presented in \Cref{tab:alt-decomp-gggg-resource}.

\section{Results and Discussion}
\label{sec:results}
\noindent
In this work, we simulated pure SU(2) gauge fields in the total angular momentum basis, where each link is encoded by a qudit.
We decomposed the Trotterized evolution of the plaquette operator for arbitrary gauge field truncation levels into elementary qudit gates and detailed a resource tradeoff between gate counts and dimensionality of the auxiliary qudit.
The precise resource counts of these decompositions are presented in \Cref{tab:gggg-decomp-resource}, \Cref{tab:qudit-decomp-resource}, and \Cref{tab:alt-decomp-gggg-resource}.
In particular, when the auxiliary qudit is a qu8it, the gate counts scale linearly with the number of non-zero plaquette operator transition matrix elements, which we enumerate precisely in \Cref{eq:control-sector} for the SU(2) gauge group.
These are, to the best of our knowledge, the first quantum resource estimates at the elementary gate level for the evolution of the SU(2) plaquette operator in the scalable angular momentum representation.
We present these gate-level resource estimates as an upper bound to be improved through future community investigations.

In the provided code~\cite{qlgt-code}, we automatically generate the SU(2) plaquette operator GVC for arbitrary truncation levels.
Then, the current implementation, specialized to qutrits, compiles circuits directly from the qutrit plaquette operator GVC. Our code is extendable to qudits with appropriate enhancement of PennyLane (i.e., addition of qudit devices and operations).
The provided code also implements the decompositions of {\UCDBT} and {\CCDBT} for arbitrary $d$.
These decompositions, along with {\FredkinTrick}, are inspired by optimizing the presented LGT simulation, but are useful in procedures common to most quantum computing applications.
Quantifying the improvements afforded by our decompositions to general qudit compilation methods such as state preparation and unitary synthesis is the subject of future work.

We assemble the compiled qutrit circuits to simulate, in PennyLane, the real-time dynamics of qutrit-digitized SU(2) gauge fields on a cube,
verifying the correctness of the circuitry and resource estimates.
In assembling the circuitry, we utilize parallelization opportunities that arise in higher-dimensional lattice systems.

Our decompositions yield a quantum circuit that returns to the physical subspace $\mathcal{P}$ after each $\Pi\Pi\Pi\Pi\mathcal{X}\mathcal{X}\mathcal{X}\mathcal{X}$ term (and predictably transforms $\mathcal{P}$ within), enabling error-mitigation techniques to be employed at a sub-Trotter-step frequency.
Such digital approaches for physical subspace verification and protection have been explored in recent work ~\cite{Stryker:2018efp,Tran:2020azk,Lamm:2020jwv,Nguyen:2021hyk,Ballini:2024qmr,Ball:2024xmw,Wauters:2024shc,Carena:2024dzu}.
Furthermore, we anticipate that compression techniques~\cite{Bravyi:2017eoo,DiMatteo:2020dhe} can be designed to improve the efficiency of the simulation while maintaining locality and thereby scalability of the LGT representation.
These techniques become increasingly important to compensate for the diminishing ratio of physical states in the computational basis as the simulation volume and gauge field truncation are increased toward the continuum.
Along with these opportunities, our qutrit cube simulation presents a compelling, physics-driven executable for near-future qutrit hardware.

The techniques presented can be readily extended to qudit compilation of the SU(3) gauge group~\cite{Ciavarella:2021nmj}, moving towards simulations of lattice quantum chromodynamics in three spatial dimensions.
These techniques also form the foundation for a software package addressing quantum compilation challenges toward the continuum of LGT simulations.

\begin{acknowledgments}
This collaboration was funded by the NSERC Alliance International Catalyst
Quantum Grant program (ALLRP 586483-23). JJ is funded in part by the NSERC CREATE
in Quantum Computing Program, grant number 543245. NK acknowledges funding in part from the NSF STAQ Program (PHY-1818914). 
ODM acknowledges funding from NSERC, the Canada Research Chairs Program, and UBC.
This work was conceived at the 2023 \emph{Quantum Computing, Quantum Simulation, Quantum Gravity and the Standard Model} workshop at the InQubator for Quantum Simulation (IQuS) hosted by the Institute for Nuclear Theory (INT). IQuS is supported by U.S. Department of Energy, Office of Science, Office of Nuclear Physics, under Award Number DOE (NP) Award DE-SC0020970 via the program on Quantum Horizons: QIS Research and Innovation for Nuclear Science, and by the Department of Physics, and the College of Arts and Sciences at the University of Washington.
\end{acknowledgments}

\onecolumngrid
\clearpage
\twocolumngrid

\appendix
\section{\label{app:truncated-plaquette-operator}Truncated Plaquette Operator}
\FloatBarrier
\noindent
In \Cref{subsec:plaq-op}, we determined all $\PiTerm \GiTerm$ with non-zero $\phi$ in our plaquette operator GVC via \Cref{eq:control-sector}.
The values of $\phi$ can be calculated using~\cite{Klco:2019evd,ARahman:2021}, 
\begin{align}
    \langle \chi_{\cdots, j^{t,b}_{\ell}, q_{\ell f}, j^{t,b}_{af}, q_{r f}, j^{t,b}_{r}, \cdots } | \hat{\Box} | \chi_{\cdots, j^{t,b}_{\ell}, q_{\ell i}, j^{t,b}_{ai}, q_{r i}, j^{t,b}_{r}, \cdots } \rangle &= \nonumber\\* &\hspace{-6cm}
      \sqrt{\dim(j_{ai}^t)\dim(j_{af}^t) \dim(j_{ai}^b)\dim(j_{af}^b) } \nonumber \\*  &\hspace{-6.5cm} \times \sqrt{\dim(q_{\ell i})\dim(q_{\ell f}) \dim(q_{r i})\dim(q_{r f})} \label{eq:angle-formula} \\*  &\hspace{-6.5cm}
     \times (-1)^{j_\ell^t + j_\ell^b + j_r^t + j_r^b + 2(j_{af}^t + j_{af}^b -q_{\ell i} -q_{r i} )} \nonumber \\* & \hspace{-7.7cm}
    \times \scalemath{0.9}{\begin{Bmatrix}
      j_{\ell}^t & j_{ai}^t & q_{\ell i} \\
      \frac{1}{2} & q_{\ell f} & j_{af}^t
    \end{Bmatrix} \begin{Bmatrix}
      j_{\ell}^b & j_{ai}^b & q_{\ell i} \\
      \frac{1}{2} & q_{\ell f} & j_{af}^b
    \end{Bmatrix} \begin{Bmatrix}
      j_{r}^t & j_{a i}^t & q_{ri} \\
      \frac{1}{2} & q_{rf} & j_{af}^t
    \end{Bmatrix} \begin{Bmatrix}
      j_{r}^b & j_{ai}^b & q_{ri} \\
      \frac{1}{2} & q_{rf} & j_{af}^b
    \end{Bmatrix}} \nonumber
\end{align}
which is written in the language of physical flux configurations.
To find the physical flux configurations associated with a $\PiTerm \GiTerm$, we begin by setting a plaquette link to hold $x$ total angular momentum flux lines, corresponding to $\Gi_{xx^\prime}$ assigned to the link.
The remaining three plaquette links are determined by Gauss's law, subject to the first link and fixed control values.
We apply $\GiTerm$ on the resulting physical flux configuration to find its partner configuration.
These physical flux configuration pairs are input into \Cref{eq:angle-formula} to obtain $\phi$.

From the programmatic construction, we provide the growth of various terms with increasing qudit dimension $d$ in \Cref{tab:qudit-plaq-op}.
As an example, we provide the ququart plaquette operator $\hat{\square}^{(\Lambda_j=3/2)}$ in \Cref{tab:ququart-plaq-op}.
Additionally, we provide the full qutrit plaquette operator, including $\phi$, in \Cref{eq:full-qutrit-plaq-op} with a reordered Hilbert space to emphasize the organization by $\GiTerm$.

\begin{table}%
    \caption{\label{tab:qudit-plaq-op}
    Scaling of various terms with qudit dimension~$d = c+1$: number of distinct $\GiTerm$ under $D_4$ symmetry (denoted as $\GiTerm_{D_4}$), total number of $\GiTerm$ and the number of $\PiTerm \GiTerm$ with non-zero $\phi$.
    }
    \begin{ruledtabular}
        \begin{tabular}{c|ccc}
        $d$ & \begin{tabular}{@{}c@{}}\# of $\GiTerm_{D_4}$ \\ {\small $\frac{c(c+1)(c^2+c+2)}{8}$} \end{tabular} & \begin{tabular}{@{}c@{}}\# of $\GiTerm$ \\ {\small $c^4$} \end{tabular} & \begin{tabular}{@{}c@{}}\# of $\PiTerm \GiTerm$ \\ {\small $\sum_{pqrs}|\mathcal{C}_{pqrs}|$}\end{tabular}\\
        \hline
        2 & 1  & 1   & 8\\
        3 & 6  & 16  & 217\\
        4 & 21 & 81  & 2346\\
        5 & 55 & 256 & 14872\\
        6 & 120 & 625 & 66950 \\
        7 & 231 & 1296 & 237981 \\
        8 & 406 & 2401 & 711828 \\
        9 & 666 & 4096 & 1866940
        \end{tabular}
    \end{ruledtabular}
\end{table}

\onecolumngrid
\vbox{
    \tiny
    \begin{alignat*}{11}
        &\qtplaq = \\
        & \Gi_{01}\Gi_{01}\Gi_{01}\Gi_{01}\Big(&& 1 && \Pi_{0}\Pi_{0}\Pi_{0}\Pi_{0} + \frac{1}{2} && \Pi_{0}\Pi_{0}\Pi_{1}\Pi_{1} - \frac{1}{2} && \Pi_{0}\Pi_{1}\Pi_{0}\Pi_{1} + \frac{1}{2} && \Pi_{0}\Pi_{1}\Pi_{1}\Pi_{0} + \frac{1}{2} && \Pi_{1}\Pi_{0}\Pi_{0}\Pi_{1} - \frac{1}{2} && \Pi_{1}\Pi_{0}\Pi_{1}\Pi_{0} + \frac{1}{2} && \Pi_{1}\Pi_{1}\Pi_{0}\Pi_{0} + \frac{1}{4} && \Pi_{1}\Pi_{1}\Pi_{1}\Pi_{1}\Big) + && \nonumber \\
        & \Gi_{01}\Gi_{01}\Gi_{01}\Gi_{12}\Big(&& \frac{\sqrt{3}}{2} && \Pi_{1}\Pi_{0}\Pi_{0}\Pi_{1} + \frac{1}{2} && \Pi_{1}\Pi_{0}\Pi_{1}\Pi_{2} - \frac{1}{2} && \Pi_{1}\Pi_{1}\Pi_{0}\Pi_{2} + \frac{\sqrt{3}}{4} && \Pi_{1}\Pi_{1}\Pi_{1}\Pi_{1} + \frac{\sqrt{3}}{3} && \Pi_{2}\Pi_{0}\Pi_{0}\Pi_{2} - \frac{1}{2} && \Pi_{2}\Pi_{0}\Pi_{1}\Pi_{1} + \frac{1}{2} && \Pi_{2}\Pi_{1}\Pi_{0}\Pi_{1} + \frac{\sqrt{3}}{6} && \Pi_{2}\Pi_{1}\Pi_{1}\Pi_{2}\Big) + && \nonumber\\
        & \Gi_{01}\Gi_{01}\Gi_{12}\Gi_{01}\Big(&& \frac{\sqrt{3}}{2} && \Pi_{0}\Pi_{0}\Pi_{1}\Pi_{1} + \frac{\sqrt{3}}{3} && \Pi_{0}\Pi_{0}\Pi_{2}\Pi_{2} - \frac{1}{2} && \Pi_{0}\Pi_{1}\Pi_{1}\Pi_{2} + \frac{1}{2} && \Pi_{0}\Pi_{1}\Pi_{2}\Pi_{1} + \frac{1}{2} && \Pi_{1}\Pi_{0}\Pi_{1}\Pi_{2} - \frac{1}{2} && \Pi_{1}\Pi_{0}\Pi_{2}\Pi_{1} + \frac{\sqrt{3}}{4} && \Pi_{1}\Pi_{1}\Pi_{1}\Pi_{1} + \frac{\sqrt{3}}{6} && \Pi_{1}\Pi_{1}\Pi_{2}\Pi_{2}\Big) + && \nonumber\\
        & \Gi_{01}\Gi_{01}\Gi_{12}\Gi_{12}\Big(&& \frac{\sqrt{3}}{2} && \Pi_{1}\Pi_{0}\Pi_{1}\Pi_{0} + \frac{\sqrt{2}}{2} && \Pi_{1}\Pi_{0}\Pi_{1}\Pi_{2} + \frac{\sqrt{3}}{6} && \Pi_{1}\Pi_{0}\Pi_{2}\Pi_{1} - \frac{1}{4} && \Pi_{1}\Pi_{1}\Pi_{1}\Pi_{1} + \frac{1}{2} && \Pi_{1}\Pi_{1}\Pi_{2}\Pi_{0} + \frac{\sqrt{6}}{6} && \Pi_{1}\Pi_{1}\Pi_{2}\Pi_{2} + \frac{\sqrt{3}}{6} && \Pi_{2}\Pi_{0}\Pi_{1}\Pi_{1} - \frac{\sqrt{3}}{3} && \Pi_{2}\Pi_{0}\Pi_{2}\Pi_{0} - && \nonumber\\
        &                              && \frac{\sqrt{2}}{3} && \Pi_{2}\Pi_{0}\Pi_{2}\Pi_{2} + \frac{1}{2} && \Pi_{2}\Pi_{1}\Pi_{1}\Pi_{0} + \frac{\sqrt{6}}{6} && \Pi_{2}\Pi_{1}\Pi_{1}\Pi_{2} + \frac{1}{6} && \Pi_{2}\Pi_{1}\Pi_{2}\Pi_{1}\Big) + && \nonumber\\
        & \Gi_{01}\Gi_{12}\Gi_{01}\Gi_{01}\Big(&& \frac{\sqrt{3}}{2} && \Pi_{0}\Pi_{1}\Pi_{1}\Pi_{0} + \frac{1}{2} && \Pi_{0}\Pi_{1}\Pi_{2}\Pi_{1} - \frac{1}{2} && \Pi_{0}\Pi_{2}\Pi_{1}\Pi_{1} + \frac{\sqrt{3}}{3} && \Pi_{0}\Pi_{2}\Pi_{2}\Pi_{0} + \frac{\sqrt{3}}{4} && \Pi_{1}\Pi_{1}\Pi_{1}\Pi_{1} - \frac{1}{2} && \Pi_{1}\Pi_{1}\Pi_{2}\Pi_{0} + \frac{1}{2} && \Pi_{1}\Pi_{2}\Pi_{1}\Pi_{0} + \frac{\sqrt{3}}{6} && \Pi_{1}\Pi_{2}\Pi_{2}\Pi_{1}\Big) + && \nonumber\\
        & \Gi_{01}\Gi_{12}\Gi_{01}\Gi_{12}\Big(&& \frac{3}{4} && \Pi_{1}\Pi_{1}\Pi_{1}\Pi_{1} + \frac{1}{2} && \Pi_{1}\Pi_{1}\Pi_{2}\Pi_{2} - \frac{1}{2} && \Pi_{1}\Pi_{2}\Pi_{1}\Pi_{2} + \frac{1}{2} && \Pi_{1}\Pi_{2}\Pi_{2}\Pi_{1} + \frac{1}{2} && \Pi_{2}\Pi_{1}\Pi_{1}\Pi_{2} - \frac{1}{2} && \Pi_{2}\Pi_{1}\Pi_{2}\Pi_{1} + \frac{1}{2} && \Pi_{2}\Pi_{2}\Pi_{1}\Pi_{1} + \frac{1}{3} && \Pi_{2}\Pi_{2}\Pi_{2}\Pi_{2}\Big) + && \nonumber\\
        & \Gi_{01}\Gi_{12}\Gi_{12}\Gi_{01}\Big(&& \frac{\sqrt{3}}{2} && \Pi_{0}\Pi_{1}\Pi_{0}\Pi_{1} + \frac{\sqrt{3}}{6} && \Pi_{0}\Pi_{1}\Pi_{1}\Pi_{2} + \frac{\sqrt{2}}{2} && \Pi_{0}\Pi_{1}\Pi_{2}\Pi_{1} - \frac{\sqrt{3}}{3} && \Pi_{0}\Pi_{2}\Pi_{0}\Pi_{2} + \frac{\sqrt{3}}{6} && \Pi_{0}\Pi_{2}\Pi_{1}\Pi_{1} - \frac{\sqrt{2}}{3} && \Pi_{0}\Pi_{2}\Pi_{2}\Pi_{2} + \frac{1}{2} && \Pi_{1}\Pi_{1}\Pi_{0}\Pi_{2} - \frac{1}{4} && \Pi_{1}\Pi_{1}\Pi_{1}\Pi_{1} + && \nonumber\\
        &                              && \frac{\sqrt{6}}{6} && \Pi_{1}\Pi_{1}\Pi_{2}\Pi_{2} + \frac{1}{2} && \Pi_{1}\Pi_{2}\Pi_{0}\Pi_{1} + \frac{1}{6} && \Pi_{1}\Pi_{2}\Pi_{1}\Pi_{2} + \frac{\sqrt{6}}{6} && \Pi_{1}\Pi_{2}\Pi_{2}\Pi_{1}\Big) + && \nonumber\\
        & \Gi_{01}\Gi_{12}\Gi_{12}\Gi_{12}\Big(&& \frac{\sqrt{3}}{2} && \Pi_{1}\Pi_{1}\Pi_{0}\Pi_{0} + \frac{\sqrt{2}}{2} && \Pi_{1}\Pi_{1}\Pi_{0}\Pi_{2} + \frac{\sqrt{3}}{12} && \Pi_{1}\Pi_{1}\Pi_{1}\Pi_{1} + \frac{\sqrt{2}}{2} && \Pi_{1}\Pi_{1}\Pi_{2}\Pi_{0} + \frac{\sqrt{3}}{3} && \Pi_{1}\Pi_{1}\Pi_{2}\Pi_{2} - \frac{\sqrt{3}}{6} && \Pi_{1}\Pi_{2}\Pi_{0}\Pi_{1} + \frac{\sqrt{3}}{6} && \Pi_{1}\Pi_{2}\Pi_{1}\Pi_{0} + \frac{\sqrt{2}}{6} && \Pi_{1}\Pi_{2}\Pi_{1}\Pi_{2} - && \nonumber\\
        &                              && \frac{\sqrt{2}}{6} && \Pi_{1}\Pi_{2}\Pi_{2}\Pi_{1} + \frac{\sqrt{3}}{6} && \Pi_{2}\Pi_{1}\Pi_{0}\Pi_{1} - \frac{\sqrt{3}}{6} && \Pi_{2}\Pi_{1}\Pi_{1}\Pi_{0} - \frac{\sqrt{2}}{6} && \Pi_{2}\Pi_{1}\Pi_{1}\Pi_{2} + \frac{\sqrt{2}}{6} && \Pi_{2}\Pi_{1}\Pi_{2}\Pi_{1} + \frac{\sqrt{3}}{3} && \Pi_{2}\Pi_{2}\Pi_{0}\Pi_{0} + \frac{\sqrt{2}}{3} && \Pi_{2}\Pi_{2}\Pi_{0}\Pi_{2} + \frac{\sqrt{3}}{18} && \Pi_{2}\Pi_{2}\Pi_{1}\Pi_{1} + && \nonumber\\
        &                              && \frac{\sqrt{2}}{3} && \Pi_{2}\Pi_{2}\Pi_{2}\Pi_{0} + \frac{2 \sqrt{3}}{9} && \Pi_{2}\Pi_{2}\Pi_{2}\Pi_{2}\Big) + && \nonumber\\
        & \Gi_{12}\Gi_{01}\Gi_{01}\Gi_{01}\Big(&& \frac{\sqrt{3}}{2} && \Pi_{1}\Pi_{1}\Pi_{0}\Pi_{0} + \frac{\sqrt{3}}{4} && \Pi_{1}\Pi_{1}\Pi_{1}\Pi_{1} - \frac{1}{2} && \Pi_{1}\Pi_{2}\Pi_{0}\Pi_{1} + \frac{1}{2} && \Pi_{1}\Pi_{2}\Pi_{1}\Pi_{0} + \frac{1}{2} && \Pi_{2}\Pi_{1}\Pi_{0}\Pi_{1} - \frac{1}{2} && \Pi_{2}\Pi_{1}\Pi_{1}\Pi_{0} + \frac{\sqrt{3}}{3} && \Pi_{2}\Pi_{2}\Pi_{0}\Pi_{0} + \frac{\sqrt{3}}{6} && \Pi_{2}\Pi_{2}\Pi_{1}\Pi_{1}\Big) + && \nonumber\\
        & \Gi_{12}\Gi_{01}\Gi_{01}\Gi_{12}\Big(&& \frac{\sqrt{3}}{2} && \Pi_{0}\Pi_{1}\Pi_{0}\Pi_{1} + \frac{1}{2} && \Pi_{0}\Pi_{1}\Pi_{1}\Pi_{2} - \frac{\sqrt{3}}{3} && \Pi_{0}\Pi_{2}\Pi_{0}\Pi_{2} + \frac{1}{2} && \Pi_{0}\Pi_{2}\Pi_{1}\Pi_{1} + \frac{\sqrt{3}}{6} && \Pi_{1}\Pi_{1}\Pi_{0}\Pi_{2} - \frac{1}{4} && \Pi_{1}\Pi_{1}\Pi_{1}\Pi_{1} + \frac{\sqrt{3}}{6} && \Pi_{1}\Pi_{2}\Pi_{0}\Pi_{1} + \frac{1}{6} && \Pi_{1}\Pi_{2}\Pi_{1}\Pi_{2} + && \nonumber\\
        &                              && \frac{\sqrt{2}}{2} && \Pi_{2}\Pi_{1}\Pi_{0}\Pi_{1} + \frac{\sqrt{6}}{6} && \Pi_{2}\Pi_{1}\Pi_{1}\Pi_{2} - \frac{\sqrt{2}}{3} && \Pi_{2}\Pi_{2}\Pi_{0}\Pi_{2} + \frac{\sqrt{6}}{6} && \Pi_{2}\Pi_{2}\Pi_{1}\Pi_{1}\Big) + && \nonumber\\
        & \Gi_{12}\Gi_{01}\Gi_{12}\Gi_{01}\Big(&& \frac{3}{4} && \Pi_{1}\Pi_{1}\Pi_{1}\Pi_{1} + \frac{1}{2} && \Pi_{1}\Pi_{1}\Pi_{2}\Pi_{2} - \frac{1}{2} && \Pi_{1}\Pi_{2}\Pi_{1}\Pi_{2} + \frac{1}{2} && \Pi_{1}\Pi_{2}\Pi_{2}\Pi_{1} + \frac{1}{2} && \Pi_{2}\Pi_{1}\Pi_{1}\Pi_{2} - \frac{1}{2} && \Pi_{2}\Pi_{1}\Pi_{2}\Pi_{1} + \frac{1}{2} && \Pi_{2}\Pi_{2}\Pi_{1}\Pi_{1} + \frac{1}{3} && \Pi_{2}\Pi_{2}\Pi_{2}\Pi_{2}\Big) + && \nonumber\\
        & \Gi_{12}\Gi_{01}\Gi_{12}\Gi_{12}\Big(&& \frac{\sqrt{3}}{2} && \Pi_{0}\Pi_{1}\Pi_{1}\Pi_{0} + \frac{\sqrt{2}}{2} && \Pi_{0}\Pi_{1}\Pi_{1}\Pi_{2} + \frac{\sqrt{3}}{6} && \Pi_{0}\Pi_{1}\Pi_{2}\Pi_{1} - \frac{\sqrt{3}}{6} && \Pi_{0}\Pi_{2}\Pi_{1}\Pi_{1} + \frac{\sqrt{3}}{3} && \Pi_{0}\Pi_{2}\Pi_{2}\Pi_{0} + \frac{\sqrt{2}}{3} && \Pi_{0}\Pi_{2}\Pi_{2}\Pi_{2} + \frac{\sqrt{3}}{12} && \Pi_{1}\Pi_{1}\Pi_{1}\Pi_{1} - \frac{\sqrt{3}}{6} && \Pi_{1}\Pi_{1}\Pi_{2}\Pi_{0} - && \nonumber\\
        &                              && \frac{\sqrt{2}}{6} && \Pi_{1}\Pi_{1}\Pi_{2}\Pi_{2} + \frac{\sqrt{3}}{6} && \Pi_{1}\Pi_{2}\Pi_{1}\Pi_{0} + \frac{\sqrt{2}}{6} && \Pi_{1}\Pi_{2}\Pi_{1}\Pi_{2} + \frac{\sqrt{3}}{18} && \Pi_{1}\Pi_{2}\Pi_{2}\Pi_{1} + \frac{\sqrt{2}}{2} && \Pi_{2}\Pi_{1}\Pi_{1}\Pi_{0} + \frac{\sqrt{3}}{3} && \Pi_{2}\Pi_{1}\Pi_{1}\Pi_{2} + \frac{\sqrt{2}}{6} && \Pi_{2}\Pi_{1}\Pi_{2}\Pi_{1} - \frac{\sqrt{2}}{6} && \Pi_{2}\Pi_{2}\Pi_{1}\Pi_{1} + && \nonumber\\
        &                              && \frac{\sqrt{2}}{3} && \Pi_{2}\Pi_{2}\Pi_{2}\Pi_{0} + \frac{2 \sqrt{3}}{9} && \Pi_{2}\Pi_{2}\Pi_{2}\Pi_{2}\Big) + && \nonumber\\
        & \Gi_{12}\Gi_{12}\Gi_{01}\Gi_{01}\Big(&& \frac{\sqrt{3}}{2} && \Pi_{1}\Pi_{0}\Pi_{1}\Pi_{0} + \frac{1}{2} && \Pi_{1}\Pi_{0}\Pi_{2}\Pi_{1} - \frac{1}{4} && \Pi_{1}\Pi_{1}\Pi_{1}\Pi_{1} + \frac{\sqrt{3}}{6} && \Pi_{1}\Pi_{1}\Pi_{2}\Pi_{0} + \frac{\sqrt{2}}{2} && \Pi_{1}\Pi_{2}\Pi_{1}\Pi_{0} + \frac{\sqrt{6}}{6} && \Pi_{1}\Pi_{2}\Pi_{2}\Pi_{1} + \frac{1}{2} && \Pi_{2}\Pi_{0}\Pi_{1}\Pi_{1} - \frac{\sqrt{3}}{3} && \Pi_{2}\Pi_{0}\Pi_{2}\Pi_{0} + && \nonumber\\
        &                              && \frac{\sqrt{3}}{6} && \Pi_{2}\Pi_{1}\Pi_{1}\Pi_{0} + \frac{1}{6} && \Pi_{2}\Pi_{1}\Pi_{2}\Pi_{1} + \frac{\sqrt{6}}{6} && \Pi_{2}\Pi_{2}\Pi_{1}\Pi_{1} - \frac{\sqrt{2}}{3} && \Pi_{2}\Pi_{2}\Pi_{2}\Pi_{0}\Big) + && \nonumber\\
        & \Gi_{12}\Gi_{12}\Gi_{01}\Gi_{12}\Big(&& \frac{\sqrt{3}}{2} && \Pi_{0}\Pi_{0}\Pi_{1}\Pi_{1} + \frac{\sqrt{3}}{3} && \Pi_{0}\Pi_{0}\Pi_{2}\Pi_{2} - \frac{\sqrt{3}}{6} && \Pi_{0}\Pi_{1}\Pi_{1}\Pi_{2} + \frac{\sqrt{3}}{6} && \Pi_{0}\Pi_{1}\Pi_{2}\Pi_{1} + \frac{\sqrt{2}}{2} && \Pi_{0}\Pi_{2}\Pi_{1}\Pi_{1} + \frac{\sqrt{2}}{3} && \Pi_{0}\Pi_{2}\Pi_{2}\Pi_{2} + \frac{\sqrt{3}}{6} && \Pi_{1}\Pi_{0}\Pi_{1}\Pi_{2} - \frac{\sqrt{3}}{6} && \Pi_{1}\Pi_{0}\Pi_{2}\Pi_{1} + && \nonumber\\
        &                              && \frac{\sqrt{3}}{12} && \Pi_{1}\Pi_{1}\Pi_{1}\Pi_{1} + \frac{\sqrt{3}}{18} && \Pi_{1}\Pi_{1}\Pi_{2}\Pi_{2} + \frac{\sqrt{2}}{6} && \Pi_{1}\Pi_{2}\Pi_{1}\Pi_{2} - \frac{\sqrt{2}}{6} && \Pi_{1}\Pi_{2}\Pi_{2}\Pi_{1} + \frac{\sqrt{2}}{2} && \Pi_{2}\Pi_{0}\Pi_{1}\Pi_{1} + \frac{\sqrt{2}}{3} && \Pi_{2}\Pi_{0}\Pi_{2}\Pi_{2} - \frac{\sqrt{2}}{6} && \Pi_{2}\Pi_{1}\Pi_{1}\Pi_{2} + \frac{\sqrt{2}}{6} && \Pi_{2}\Pi_{1}\Pi_{2}\Pi_{1} + && \nonumber\\
        &                              && \frac{\sqrt{3}}{3} && \Pi_{2}\Pi_{2}\Pi_{1}\Pi_{1} + \frac{2 \sqrt{3}}{9} && \Pi_{2}\Pi_{2}\Pi_{2}\Pi_{2}\Big) + && \nonumber\\
        & \Gi_{12}\Gi_{12}\Gi_{12}\Gi_{01}\Big(&& \frac{\sqrt{3}}{2} && \Pi_{1}\Pi_{0}\Pi_{0}\Pi_{1} + \frac{\sqrt{3}}{6} && \Pi_{1}\Pi_{0}\Pi_{1}\Pi_{2} + \frac{\sqrt{2}}{2} && \Pi_{1}\Pi_{0}\Pi_{2}\Pi_{1} - \frac{\sqrt{3}}{6} && \Pi_{1}\Pi_{1}\Pi_{0}\Pi_{2} + \frac{\sqrt{3}}{12} && \Pi_{1}\Pi_{1}\Pi_{1}\Pi_{1} - \frac{\sqrt{2}}{6} && \Pi_{1}\Pi_{1}\Pi_{2}\Pi_{2} + \frac{\sqrt{2}}{2} && \Pi_{1}\Pi_{2}\Pi_{0}\Pi_{1} + \frac{\sqrt{2}}{6} && \Pi_{1}\Pi_{2}\Pi_{1}\Pi_{2} + && \nonumber\\
        &                              && \frac{\sqrt{3}}{3} && \Pi_{1}\Pi_{2}\Pi_{2}\Pi_{1} + \frac{\sqrt{3}}{3} && \Pi_{2}\Pi_{0}\Pi_{0}\Pi_{2} - \frac{\sqrt{3}}{6} && \Pi_{2}\Pi_{0}\Pi_{1}\Pi_{1} + \frac{\sqrt{2}}{3} && \Pi_{2}\Pi_{0}\Pi_{2}\Pi_{2} + \frac{\sqrt{3}}{6} && \Pi_{2}\Pi_{1}\Pi_{0}\Pi_{1} + \frac{\sqrt{3}}{18} && \Pi_{2}\Pi_{1}\Pi_{1}\Pi_{2} + \frac{\sqrt{2}}{6} && \Pi_{2}\Pi_{1}\Pi_{2}\Pi_{1} + \frac{\sqrt{2}}{3} && \Pi_{2}\Pi_{2}\Pi_{0}\Pi_{2} - && \nonumber\\
        &                              && \frac{\sqrt{2}}{6} && \Pi_{2}\Pi_{2}\Pi_{1}\Pi_{1} + \frac{2 \sqrt{3}}{9} && \Pi_{2}\Pi_{2}\Pi_{2}\Pi_{2}\Big) + && \nonumber\\
        & \Gi_{12}\Gi_{12}\Gi_{12}\Gi_{12}\Big(&& 1 && \Pi_{0}\Pi_{0}\Pi_{0}\Pi_{0} + \frac{\sqrt{6}}{3} && \Pi_{0}\Pi_{0}\Pi_{0}\Pi_{2} + \frac{1}{6} && \Pi_{0}\Pi_{0}\Pi_{1}\Pi_{1} + \frac{\sqrt{6}}{3} && \Pi_{0}\Pi_{0}\Pi_{2}\Pi_{0} + \frac{2}{3} && \Pi_{0}\Pi_{0}\Pi_{2}\Pi_{2} - \frac{1}{6} && \Pi_{0}\Pi_{1}\Pi_{0}\Pi_{1} + \frac{1}{6} && \Pi_{0}\Pi_{1}\Pi_{1}\Pi_{0} + \frac{\sqrt{6}}{18} && \Pi_{0}\Pi_{1}\Pi_{1}\Pi_{2} - && \nonumber\\
        &                              && \frac{\sqrt{6}}{18} && \Pi_{0}\Pi_{1}\Pi_{2}\Pi_{1} + \frac{\sqrt{6}}{3} && \Pi_{0}\Pi_{2}\Pi_{0}\Pi_{0} + \frac{2}{3} && \Pi_{0}\Pi_{2}\Pi_{0}\Pi_{2} + \frac{\sqrt{6}}{18} && \Pi_{0}\Pi_{2}\Pi_{1}\Pi_{1} + \frac{2}{3} && \Pi_{0}\Pi_{2}\Pi_{2}\Pi_{0} + \frac{2 \sqrt{6}}{9} && \Pi_{0}\Pi_{2}\Pi_{2}\Pi_{2} + \frac{1}{6} && \Pi_{1}\Pi_{0}\Pi_{0}\Pi_{1} - \frac{1}{6} && \Pi_{1}\Pi_{0}\Pi_{1}\Pi_{0} - && \nonumber\\
        &                              && \frac{\sqrt{6}}{18} && \Pi_{1}\Pi_{0}\Pi_{1}\Pi_{2} + \frac{\sqrt{6}}{18} && \Pi_{1}\Pi_{0}\Pi_{2}\Pi_{1} + \frac{1}{6} && \Pi_{1}\Pi_{1}\Pi_{0}\Pi_{0} + \frac{\sqrt{6}}{18} && \Pi_{1}\Pi_{1}\Pi_{0}\Pi_{2} + \frac{1}{36} && \Pi_{1}\Pi_{1}\Pi_{1}\Pi_{1} + \frac{\sqrt{6}}{18} && \Pi_{1}\Pi_{1}\Pi_{2}\Pi_{0} + \frac{1}{9} && \Pi_{1}\Pi_{1}\Pi_{2}\Pi_{2} + \frac{\sqrt{6}}{18} && \Pi_{1}\Pi_{2}\Pi_{0}\Pi_{1} - && \nonumber\\
        &                              && \frac{\sqrt{6}}{18} && \Pi_{1}\Pi_{2}\Pi_{1}\Pi_{0} - \frac{1}{9} && \Pi_{1}\Pi_{2}\Pi_{1}\Pi_{2} + \frac{1}{9} && \Pi_{1}\Pi_{2}\Pi_{2}\Pi_{1} + \frac{\sqrt{6}}{3} && \Pi_{2}\Pi_{0}\Pi_{0}\Pi_{0} + \frac{2}{3} && \Pi_{2}\Pi_{0}\Pi_{0}\Pi_{2} + \frac{\sqrt{6}}{18} && \Pi_{2}\Pi_{0}\Pi_{1}\Pi_{1} + \frac{2}{3} && \Pi_{2}\Pi_{0}\Pi_{2}\Pi_{0} + \frac{2 \sqrt{6}}{9} && \Pi_{2}\Pi_{0}\Pi_{2}\Pi_{2} - && \nonumber\\
        &                              && \frac{\sqrt{6}}{18} && \Pi_{2}\Pi_{1}\Pi_{0}\Pi_{1} + \frac{\sqrt{6}}{18} && \Pi_{2}\Pi_{1}\Pi_{1}\Pi_{0} + \frac{1}{9} && \Pi_{2}\Pi_{1}\Pi_{1}\Pi_{2} - \frac{1}{9} && \Pi_{2}\Pi_{1}\Pi_{2}\Pi_{1} + \frac{2}{3} && \Pi_{2}\Pi_{2}\Pi_{0}\Pi_{0} + \frac{2 \sqrt{6}}{9} && \Pi_{2}\Pi_{2}\Pi_{0}\Pi_{2} + \frac{1}{9} && \Pi_{2}\Pi_{2}\Pi_{1}\Pi_{1} + \frac{2 \sqrt{6}}{9} && \Pi_{2}\Pi_{2}\Pi_{2}\Pi_{0} + && \nonumber\\
        &                              && \frac{4}{9} && \Pi_{2}\Pi_{2}\Pi_{2}\Pi_{2}\Big)
    \end{alignat*}
    \normalsize
    }
\begin{equation}
    \label{eq:full-qutrit-plaq-op}
\end{equation}
\clearpage
\twocolumngrid

\begin{table}%
    \caption{\label{tab:ququart-plaq-op}
        The $\PiTerm \GiTerm$ with non-zero $\phi$ values in the ququart plaquette operator.
        Under $D_4$ symmetry, there are $21$ representative ways to assign $\Gi \in \{\Gi_{01}, \Gi_{12}, \Gi_{23}\}$ to each $\Gi$ in $\GiTerm$.
        These representative classes are paired with $\PiTerm$ from the set $\mathcal{C}_{pqrs}$ defined in \Cref{eq:control-sector}.
        The control words ``$ijkl$'' are shown.
        We write {\es} ({\os}) to denote cases where both the $0$ and $2$ ($1$ and $3$) control values are present.
        To obtain the complete set of $81$ $\GiTerm$, the $pqrs$ indices are permuted using $D_4$ transformations;  
        \enquote{order} indicates the number of permutations leading to unique $\GiTerm$ assignments.
        The ququart plaquette operator has $2346$ distinct $\PiTerm \GiTerm$ with non-zero $\phi$.
    }
    \begin{ruledtabular}
        \begin{tabular}{cc|cc}
        \multicolumn{2}{c}{$\Gi_{p p'} \Gi_{q q'} \Gi_{r r'} \Gi_{ss'}$} & \multicolumn{2}{c}{$\Pi_i\Pi_j\Pi_k\Pi_l$} \\
        $pqrs$ & order & $ijkl$ & $|\mathcal{C}_{pqrs}|$\\ 
        \hline
        $0000$ & 1 &
            \makecell[c]{
                $\begin{array}{cccc}
                    0000 & 0011 & 0101 & 0110 \\
                    1001 & 1010 & 1100 & 1111 
                \end{array}$
            } & 8  \\
        \hline
        $0001$ & 4 &
            \makecell[c]{
                $\begin{array}{cccc}
                    1001 & 1012 & 1102 & 1111 \\
                    2002 & 2011 & 2101 & 2112 
                \end{array}$
            } & 8  \\
        \hline
        $0002$ & 4 &
            \makecell[c]{
                $\begin{array}{cccc}
                    2002 & 2013 & 2103 & 2112 \\
                    3003 & 3012 & 3102 & 3113 
                \end{array}$
            } & 8  \\
        \hline
        $0011$ & 4 &
            \makecell[c]{
                $\begin{array}{cccc}
                    101\es & 102\os & 111\os & 112\es \\
                    201\os & 202\es & 211\es & 212\os
                \end{array}$
            } & 16  \\
        \hline
        $0012$ & 8 &
            \makecell[c]{
                $\begin{array}{cccc}
                    201\os & 2022 & 2112 & 212\os \\
                    3012 & 302\os & 311\os & 3122 
                \end{array}$
            } & 12  \\
        \hline
        $0022$ & 4 &
            \makecell[c]{
                $\begin{array}{cccc}
                    202\es & 203\os & 212\os & 213\es \\
                    302\os & 303\es & 312\es & 313\os
                \end{array}$
            } & 16  \\
        \hline
        $0101$ & 2 &
            \makecell[c]{
                $\begin{array}{cccc}
                    1111 & 1122 & 1212 & 1221 \\
                    2112 & 2121 & 2211 & 2222 
                \end{array}$
            } & 8  \\
        \hline
        $0102$ & 4 &
            \makecell[c]{
                $\begin{array}{cccc}
                    2112 & 2123 & 2213 & 2222 \\
                    3113 & 3122 & 3212 & 3223 
                \end{array}$
            } & 8  \\
        \hline
        $0111$ & 4 &
            \makecell[c]{
                $\begin{array}{cccc}
                    11\es\es & 11\os\os & 12\es\os & 12\os\es \\
                    21\es\os & 21\os\es & 22\es\es & 22\os\os
                \end{array}$
            } & 32  \\
        \hline
        $0112$ & 8 &
            \makecell[c]{
                $\begin{array}{cccc}
                    21\es\os & 21\os2 & 22\es2 & 22\os\os \\
                    31\es2 & 31\os\os & 32\es\os & 32\os2
                \end{array}$
            } & 24  \\
        \hline
        $0121$ & 4 &
            \makecell[c]{
                $\begin{array}{cccc}
                    11\os\os & 1122 & 12\os2 & 122\os \\
                    21\os2 & 212\os & 22\os\os & 2222 
                \end{array}$
            } & 18  \\
        \hline
        $0122$ & 8 &
            \makecell[c]{
                $\begin{array}{cccc}
                    21\os\es & 212\os & 22\os\os & 222\es \\
                    31\os\os & 312\es & 32\os\es & 322\os 
                \end{array}$
            } & 24  \\
        \hline
        $0202$ & 2 &
            \makecell[c]{
                $\begin{array}{cccc}
                    2222 & 2233 & 2323 & 2332 \\
                    3223 & 3232 & 3322 & 3333 
                \end{array}$
            } & 8  \\
        \hline
        $0212$ & 4 &
            \makecell[c]{
                $\begin{array}{cccc}
                    22\os\os & 2222 & 23\os2 & 232\os \\
                    32\os2 & 322\os & 33\os\os & 3322
                \end{array}$
            } & 18  \\
        \hline
        $0222$ & 4 &
            \makecell[c]{
                $\begin{array}{cccc}
                    22\es\es & 22\os\os & 23\es\os & 23\os\es \\
                    32\es\os & 32\os\es & 33\es\es & 33\os\os
                \end{array}$
            } & 32  \\
        \hline
        $1111$ & 1 &
            \makecell[c]{
                $\begin{array}{cccc}
                    \es\es\es\es & \es\es\os\os & \es\os\es\os & \es\os\os\es \\
                    \os\es\es\os & \os\es\os\es & \os\os\es\es & \os\os\os\os
                \end{array}$
            } & 128  \\
        \hline
        $1112$ & 4 &
            \makecell[c]{
                $\begin{array}{cccc}
                    \os\es\es\os & \os\es\os2 & \os\os\es2 & \os\os\os\os \\
                    2\es\es2 & 2\es\os\os & 2\os\es\os & 2\os\os2
                \end{array}$
            } & 72  \\
        \hline
        $1122$ & 4 &
            \makecell[c]{
                $\begin{array}{cccc}
                    \os\es\os\es & \os\es2\os & \os\os\os\os & \os\os2\es \\
                    2\es\os\os & 2\es2\es & 2\os\os\es & 2\os2\os 
                \end{array}$
            } & 72  \\
        \hline
        $1212$ & 2 &
            \makecell[c]{
                $\begin{array}{cccc}
                    \os\os\os\os & \os\os22 & \os2\os2 & \os22\os \\
                    2\os\os2 & 2\os2\os & 22\os\os & 2222
                \end{array}$
            } & 41  \\
        \hline
        $1222$ & 4 &
            \makecell[c]{
                $\begin{array}{cccc}
                    \os\os\es\es & \os\os\os\os & \os2\es\os &  \os2\os\es \\
                    2\os\es\os & 2\os\os\es & 22\es\es & 22\os\os
                \end{array}$
            } & 72  \\
        \hline
        $2222$ & 1 &
            \makecell[c]{
                $\begin{array}{cccc}
                    \es\es\es\es & \es\es\os\os & \es\os\es\os & \es\os\os\es \\
                    \os\es\es\os & \os\es\os\es & \os\os\es\es & \os\os\os\os
                \end{array}$
            } & 128  \\
        \hline
        \end{tabular}
    \end{ruledtabular}
\end{table}

\section{{\UCDBT}}
\label{app:ucdbt}
\begin{figure*}
    \includegraphics[width=\textwidth]{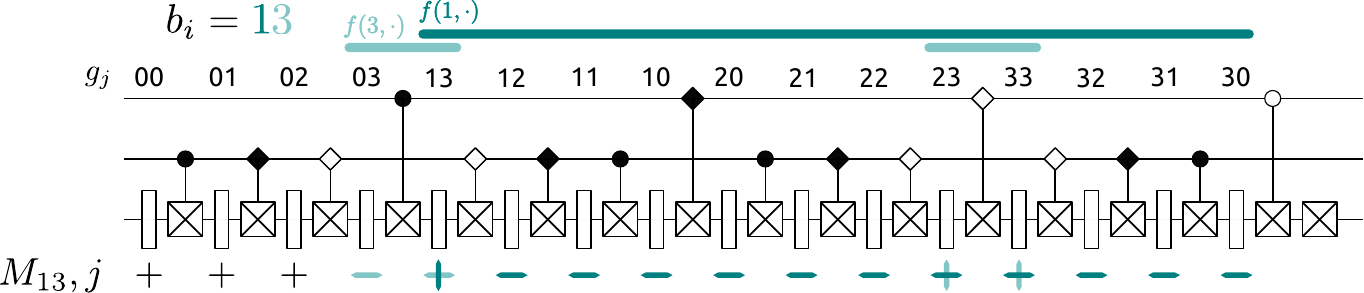}%
    \caption{\label{fig:ququart-beta-trick} 
      Circuit decomposition of $\ucnot{2}{4}{a}$, a {\UCDBT} subroutine for $d = 4$ and $k = 2$.
      An open diamond represents control on $\ket{3}$.
      In the angle calculation, $\bm{\theta} = M \bm{\beta}$, all entries of $M$ are $\pm 1$, and the sign of the element $M_{ij}$ is determined by whether an odd or even number of indices $l \in \mathbb{Z}_{k}$ satisfy the condition $0 < b_l^{(i)} \leq g_l^{(j)}$ as per \Cref{eq:qudit-beta-matrix}.
      The depicted example is for the row $M_{13}$, where $13 =
      1 \cdot 4^{1}+ 3 \cdot 4^{0} = 7$ in quaternary notation.
      To finish the circuit construction, we append three GCX gates controlled
      on $\ket{1}$, $\ket{2}$ and $\ket{3}$ to the top control ququart, and use
      the optimization in \Cref{fig:gcx-opt} to reduce this to one GCX gate
      controlled on $\ket{0}$, and an $X_{ij}$ gate.
    }
\end{figure*}
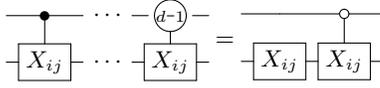
\begin{figure}
    \[
        \begin{gathered}
            \Qcircuit @C=.5em @R=.5em {
                & \tctrlo{1}    & \qw & & \cdots & & & \mctrl{1}{d \minus 1} & \qw\\ 
                & \gate{X_{ij}} & \qw & & \cdots & & & \gate{X_{ij}} & \qw
            }
        \end{gathered}
        =
        \begin{gathered}
            \Qcircuit @C=.5em @R=1.em {
                & \qw           & \tctrlz{1}    & \qw \\
                & \gate{X_{ij}} & \gate{X_{ij}} & \qw
            }   
        \end{gathered}
    \]
    \caption{\label{fig:gcx-opt}
    Applying an $X_{ij}$ gate controlled on $\{1, \cdots, d-1\}$, is the same as not controlling on  $\{0\}$.
    }
\end{figure}
\noindent
The complete resource count for $\ucnot{k}{d}{a}$ is
\begin{align*}
    \# \text{ GCXs} = & \begin{cases}
        d^k & \text{if } d \bmod 2 = 0,\\
        d^k + (k-1) & \text{otherwise},
    \end{cases}\\
    \#\ R_a = &\ d^k,\\
    \#\ X = & \begin{cases}
        1 & \text{if } d > 2 \text{ and } d \bmod 2 = 0,\\
        1 & \text{if } d \bmod 2 = 1 \text{ and } k \bmod 2 = 1,\\
        0 & \text{otherwise}.
    \end{cases}
\end{align*}
The dependence on $d$ and $k$ is due to the varying numbers of GCX gates that
must be appended to the circuit to ensure no net $R_{\mathcal{X}}$ rotation.
For odd $d$, we append $d-1$ GCX gates with control values from $1$ to $d-1$ to each control qudit, which reduces to a single GCX gate and an $X_{ij}$ gate using the optimization in \Cref{fig:gcx-opt}.
When $k$ is even, all $X_{ij}$ gates cancel, otherwise a single one remains.
For even $d$, we append $d-1$ GCX gates to the top control qudit and apply the same optimization when $d>2$.

In \Cref{fig:ququart-beta-trick}, we show the circuit decomposition of $\ucnot{2}{4}{a}$, which corresponds to uniform-control over two ququarts ($d=4, k=2$).
In the same figure, we also show the diagrammatic construction of a row in $M$, which uniquely combines $\beta$s to yield the $\theta$ corresponding to the control word indexing the row.

\begin{figure}
    \includegraphics[width=\columnwidth]{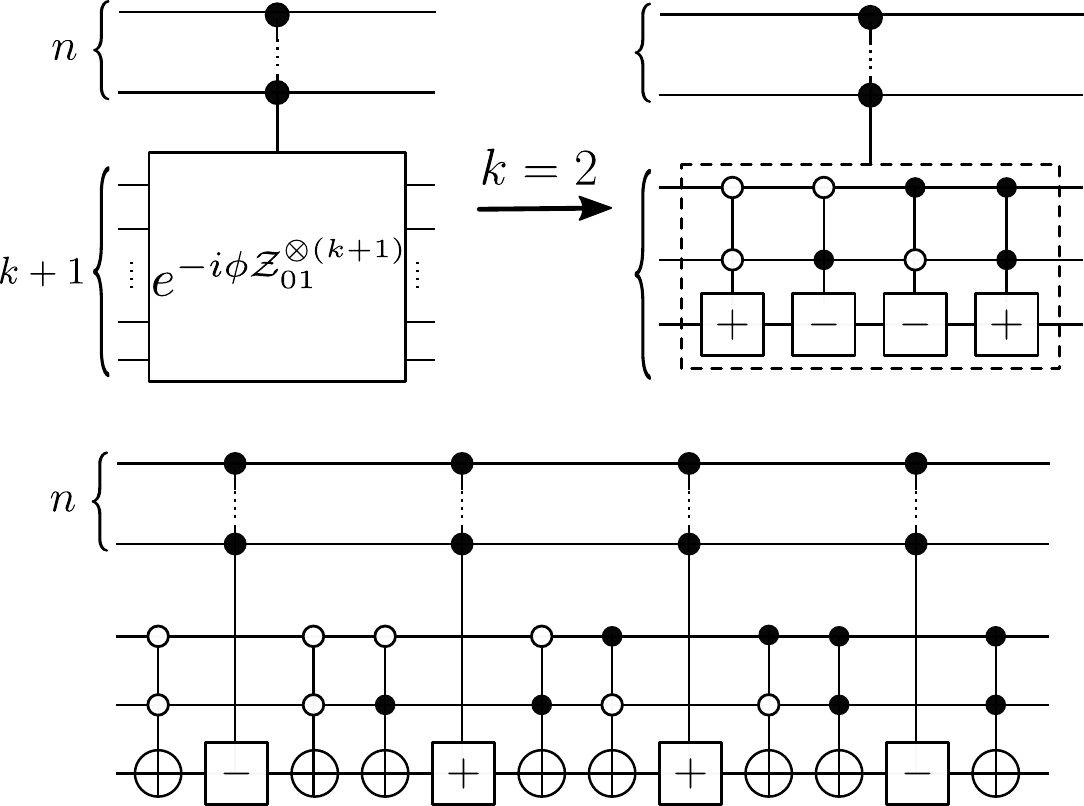}
    \caption{\label{fig:alt-angle-decomp}
    An alternate decomposition to \Cref{fig:or_fredkin_trick} of $\exp(-i \phi \Pi^n \mathcal{Z}_{01}^{\otimes k + 1})$ for arbitrary $n$ and $k$, illustrated for $k=2$. 
    Since $\mathcal{Z}_{01}^{\otimes k+1}$ contains an equal amount of positive and negative angles, we decompose controlled rotation gates in pairs where the angles $\theta$ have equal magnitude and opposite sign.
    The resulting circuit contains $2^k$ $n$-controlled rotation gates.
    }
\end{figure}

\section{Subspace Verification Circuits}
\label{app:fredkin-trick}
\noindent
The gating subcircuit accompanying a subroutine verifies the input is in the affected subspace of the subroutine and writes the result of the verification to an auxiliary qudit. 
\begin{figure}
    \includegraphics[width=\columnwidth]{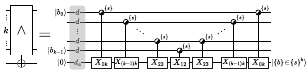}
    \caption{\label{fig:input-verification}
    The input verification circuit for qudits with ${d_a > 3}$ checks whether the input $\{b\}$ is in the $\{s\}^k$ subspace.
    The half-filled circles represent $|s|$ adjacent GCX gates (with control values spanning $\{s \}$) applying the same $X_{ij}$ unitary.
    This circuit construction supports $k$ input qudits as long as the auxiliary qudit has dimension $d_a > k$.
    The circuit uses $|s|(2k - 1)$ GCX (and has the same depth).
    Taking the qubit subspace as an example, $\{s\} = \{0, 1\}$, each circuit element becomes two GCX gates, one controlled on $\ket{0}$ and the other on $\ket{1}$.
    The circuit writes $\ket{1}$ to the auxiliary qudit if the input $\{b\}$ is in the chosen subspace of $\mathbb{Z}_2^k$.
    For the plaquette operator with $|s|=2$ and $k=3$, the circuit uses 10 GCX gates and has depth 10.
    }
\end{figure}
For a circuit with an auxiliary qudit with dimension $d_a > 3$, we can verify the inputs using the $\land$ gate in \Cref{fig:input-verification}.

In some cases, it is easier to verify instead that inputs are outside the subroutine's subspace. 
An example is verifying inputs to qubit subroutines in qutrit circuits.
\begin{figure}
    \includegraphics[width=\columnwidth]{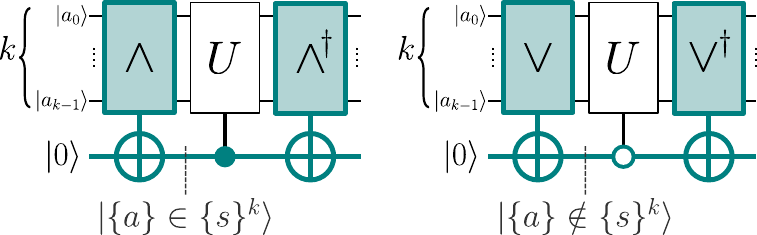}
    \caption{\label{fig:de-morgan}
    The $\land$ and $\lor$ version of the gating subcircuit.
    De Morgan's law, $(A \land B) = \neg(\neg A \lor \neg B)$, allows us to implement an $\land$ gate using an $X_{ij}$ and an $\lor$ gate.
    In the context of the gating subcircuit: the two $X_{01}$ gates flip the control value from 1 to 0, the $\land$ gates become $\lor$ gates, 
    and the inputs $a$ satisfying the Boolean condition are inverted from $\{a\} \in \{s\}$ to $\{a\} \notin \{s\}$ where $\{s\}$ is the subroutine's subspace.
    }
\end{figure}
This version of the verification is done using $\lor$ gates (through application of De Morgan's law shown in \Cref{fig:de-morgan}).

We provide the qudit $\lor$ gate construction for arbitrary input size in 
\begin{figure*}
\includegraphics[width=\textwidth]{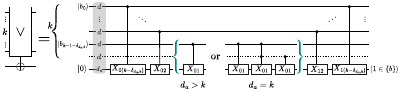}
\caption{
    \label{fig:qudit-lor-gates}
    The qudit $\lor$ gate writes $\ket{1}$ to the auxiliary qudit if any of the $k$ input qudits is in state $\ket{1}$.
    The $\lor$ gate can be configured to detect states other than $\ket{1}$ by changing the control value of the GCX gates.
    When $d_a > k$, the circuit above features the left completion and uses $2k-1$ GCX gates and depth.  
    When $d_a = k$, the right completion is used, utilizing $4k-5$ GCX gates and depth by cancelling
    $2$ GCX gates corresponding to the leftmost GCX gate in the qudit Toffoli gate (as constructed in \Cref{fig:qudit-toffoli}).
    As an example, the two-input qudit $\lor$ gate (left completion) uses 3 GCX gates, and has depth 3.
}
\end{figure*}
\begin{figure}
    \[
    \begin{gathered}
        \Qcircuit @C=.5em @R=.8em {
            & \ctrl{1} & \qw \\
            & \ctrl{1} & \qw \\
            & \gate{X_{01}} & \qw
        }
    \end{gathered}
    \hspace{.5em}
    {\displaystyle = }
    \hspace{.5em}
    \begin{gathered}
        \Qcircuit @C=.0em @R=.2em {
            & \ctrl{2} & \ctrl{1} & \cds{1}{\cdots} & \ctrl{1} & \qw & \ctrl{1} & \cds{1}{\cdots} & \ctrl{1} & \qw & \qw \\ 
            & \qw & \gate{X_{02}} & \qw & \gate{X_{0m}} & \ctrlo{1} & \gate{X_{02}} & \qw & \gate{X_{0m}} & \ctrlo{1} & \qw \\ 
            & \gate{X_{01}} & \qw & \qw & \qw & \gate{X_{01}} & \qw & \qw & \qw & \gate{X_{01}} & \qw
        }
    \end{gathered}
    \]
    \caption{\label{fig:qudit-toffoli}
    The qudit Toffoli gate, where $m=d-1$.
    The circuit uses $2d-1$ GCX gates and depth.
    }
\end{figure}
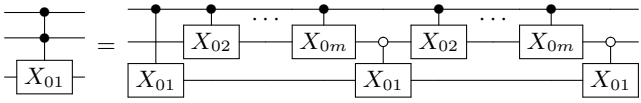
\Cref{fig:qudit-lor-gates}, subject to $d_a\geq k$.
\begin{figure}
    \[
        \hspace{-5em}
        \begin{gathered}
            \includegraphics{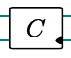}  
        \end{gathered}
        \hspace{.5em}
        {\displaystyle =}
        \hspace{1.5em}
        \begin{gathered}
            \Qcircuit @C=.5em @R=.7em {
                & \lstick{\ket{a}} & \gate{X_{12}} & \gate{X_{12}} & \tctrlt{1} & \qw & \rstick{\ket{\text{dirty}}}\\ 
                & \lstick{\ket{b}} & \qw & \tctrlo{-1} & \gate{X_{01}} & \qw & \rstick{\ket{1\in \{a, b\}}}
            }
        \end{gathered}
    \]
    \[
        \hspace{-6. em}
        \Qcircuit @C=.5em @R=.9em {
            & \lstick{\ket{e}} & \tctrlo{2} \gategroup{1}{3}{3}{5}{.3em}{--} & \qw & \tctrlo{2} & \qw & \qw & \qw & \qw & \rstick{\ket{e}} \\ 
            & \lstick{\ket{f}} & \qw & \tctrlo{1} & \qw & \qw & \qw & \qw & \qw & \rstick{\ket{f}} \\ 
            & \lstick{\ket{0}} & \gate{X_{02}} & \gate{X_{01}} & \gate{X_{12}} & \gate{X_{12}} \gategroup{3}{6}{6}{8}{.3em}{--} & \gate{X_{12}} & \tctrlt{3} & \qw & \rstick{\ket{\text{dirty}}} \\ 
            & \lstick{\ket{g}} & \tctrlo{2} \gategroup{4}{3}{6}{5}{.3em}{--} & \qw & \tctrlo{2} & \qw & \qw & \qw & \qw & \rstick{\ket{g}} \\ 
            & \lstick{\ket{h}} & \qw & \tctrlo{1} & \qw & \qw & \qw & \qw & \qw & \rstick{\ket{h}} \\ 
            & \lstick{\ket{0}} & \gate{X_{02}} & \gate{X_{01}} & \gate{X_{12}} & \qw & \tctrlo{-3} & \gate{X_{01}} & \qw & \rstick{\ket{1\in\{e,f,g,h\}}}
        }
    \]
    \caption{\label{fig:wide-qudit-lor-gates}
    (top) We construct larger $\lor$ gates from smaller $\lor$ gates using the $C$ (combine) gate to combine $\lor$ gate outputs.
    Inputs to the $C$ gate are assumed to be in the subspace $\mathbb{Z}^2_2$. The result is written on the qudit indicated by the notch.
    This construction is useful when $d_a < k$, as those cases are not addressed by the circuits in \Cref{fig:qudit-lor-gates}.
    (bottom) For example, the four-input qutrit $\lor$ gate is composed of two two-input qutrit $\lor$ gates and a $C$ gate,
    amounting to a total resource count of two auxiliary qutrits, $8$ GCX gates, an $X_{12}$ gate, and depth $6$.
    This gate is used in the fine-grained SU(2) decomposition presented in \Cref{fig:gggg-decomp-alt}.
    }
\end{figure}
When $d_a<k$, the qudit $\lor$ gate requires additional auxiliary qudits as shown in \Cref{fig:wide-qudit-lor-gates}.
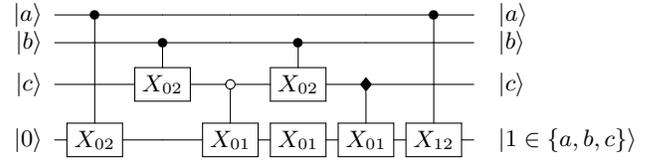
\begin{figure}
    \[
        \hspace{-5em}
        \Qcircuit @C=.5em @R=.9em {
            & \lstick{\ket{a}} & \ctrl{3} & \qw & \qw & \qw & \qw & \ctrl{3} & \qw & \rstick{\ket{a}} \\ 
            & \lstick{\ket{b}} & \qw & \ctrl{1} & \qw & \ctrl{1} & \qw & \qw & \qw & \rstick{\ket{b}} \\ 
            & \lstick{\ket{c}} & \qw & \gate{X_{02}} & \tctrlz{1} & \gate{X_{02}} & \tctrlt{1} & \qw & \qw & \rstick{\ket{c}} \\ 
            & \lstick{\ket{0}} & \gate{X_{02}} & \qw & \gate{X_{01}} & \gate{X_{01}} & \gate{X_{01}} & \gate{X_{12}} & \qw & \rstick{\ket{1 \in \{a, b, c\}}}
        }
    \]
    \caption{\label{fig:three-input-lor}
    The three-input qutrit $\lor$ gate, a special case of the three-input qudit $\lor$ gate shown in \Cref{fig:qudit-lor-gates},
    where an additional GCX gate is saved by applying the optimization in \Cref{fig:gcx-opt} on the pair of GCX gates corresponding to the rightmost GCX gate in the qutrit Toffoli gate (\Cref{fig:qudit-toffoli}).
    This three-input qutrit $\lor$ gate uses $6$ GCX gates, an $X_{01}$ gate and has depth $6$.
    }
\end{figure}
An optimized version of the qudit $\lor$ gate, the three-input qutrit $\lor$ gate (\Cref{fig:three-input-lor}), is used to synthesize the evolution of the qutrit plaquette operator.

\section{The $d$-ary and Gray Sequencers}
\label{app:ccdbt}
\noindent
The following details our decomposition of $\ccnot{k}{d}{a}$, following the intuition described in \Cref{subsec:ccdbt}.
Two algorithms, the $d$-ary sequencer and the Gray sequencer, are used to generate $\tilde{b}^\prime$ and $\tilde{g}^\prime$ respectively.
They are then normalized to yield the final sequences $b^{\prime}$ and $g^{\prime}$.

We use 
\begin{figure}
    \includegraphics[width=\columnwidth]{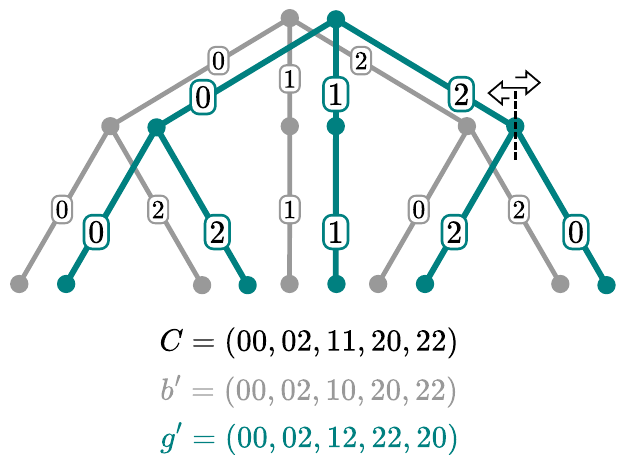}
    \[
        \Qcircuit @C=.5em @R=.7em {
            & \qw & \qw & \qw & \tctrlo{2} & \qw & \tctrlt{2} & \qw & \qw & \qw & \tctrlz{2} & \qw & \qw \\ 
            & \qw & \tctrlt{1} & \qw & \qw & \qw & \qw & \qw & \tctrlt{1} & \qw & \qw & \qw & \qw \\ 
            & \gate{\beta_0} & \targ & \gate{\beta_1} & \targ & \gate{\beta_2} & \targ & \gate{\beta_3} & \targ & \gate{\beta_4} & \targ & \targ & \qw
        }
    \]
    \caption{\label{fig:seq-tree}
        The $d$-ary (light grey) and Gray (teal) trees corresponding to $\ctrlseq = (00,02,11,20,22)$ before branch relabelling.
        Notice the reflected branches of the Gray tree at the marked position, which corresponds to reflecting branches at every second parent node with multiple child nodes.
        The Gray sequencer relabels branch ``1" which is the sole child of its parent branch to the label of the nearest left branch, which is ``2".
        Therefore, the Gray sequence is $g^\prime = (00,02,12,22,20)$. 
        In tandem, the $d$-ary sequencer relabels the corresponding branch to ``0", and so the $d$-ary sequence is $b^\prime = (00,02,10,20,22)$.
        Since $\ctrlseq$ starts with the $00$ control word, the sequence normalizer is not needed in this instance.
    }
\end{figure}\Cref{fig:seq-tree} to aid in describing the sequencers. 
The $d$-ary sequencer generates a tree, the $d$-ary tree, directly from $\ctrlseq$, where the branches are ordered by increasing digit values.
The Gray sequencer generates a tree, the Gray tree, through modified breadth-first traversal of the $d$-ary tree.
While traversing each layer, the Gray sequencer \emph{reflects} branches at every second parent node that has multiple child nodes.
Parent nodes with a single child node adopt the label of the nearest branch on the left (on the same layer).
In tandem, the $d$-ary sequencer relabels the corresponding branch in the $d$-ary tree to $0$.
The unnormalized sequences, $\tilde{b}^{\prime}$ and $\tilde{g}^{\prime}$, correspond to the labels of the path taken from the root node to the leaf nodes 
in the $d$-ary and Gray trees, respectively.

Finally, the sequence normalizer replaces the first control words of $\tilde{b}^{\prime}$ and $\tilde{g}^{\prime}$ with the $\{0\}^k$ control word by
making appropriate changes to the subsequent control words, yielding $b^{\prime}$ and $g^{\prime}$.
This final step is not needed for proper circuit functionality. Rather, it avoids GCX gates that would otherwise have been placed before the first rotation gate.

\section{From Singular to Invertible $M$}
\label{app:corr-82-ccdbt}
\noindent
In the absence of the hypothesized decomposition of {\CCDBT}, we show how to manually correct a non-invertible $M$ generated from \Cref{eq:control-sequence} by inserting GCX gates.
The matrix $M$ has dimensions $82 \times 82$ (since $|\ctrlseq| = 82$), but the rank of $M$ generated using the systematic techniques of \Cref{subsec:ucdbt} is only 76.
The manual corrections are shown for this specific case, although the general strategy can be broadly applied.

To increase the rank of $M$, we conjugate a subset of the rotation gates with GCX gates, which effects sign flips on particular matrix elements.
For example, suppose we conjugate the rotation gate corresponding to column $3$ of $M$ in \Cref{fig:ququart-beta-trick} (rotation gate at $g_3=03$),
and control the GCX pair on the bottom control ququart with control value $1$.
The modified $M$ has the sign flipped for the matrix elements corresponding to the column indexed by $g_3=03$ in the Gray code and the row indexed by $b_i=\ast 1$,
i.e., all rows indexed by a $d$-ary string that ends in $1$.

We apply a search-based technique to find appropriate corrections to a non-invertible $M$.
Beginning with the lowest indexed column in the nullspace of $M$, 
we perform a brute force search for the control qudit and control value, applying the correction and updating the nullspace if the rank of $M$ increases.
If no suitable controlled operation is found, we increment the nullspace column index and repeat.
We do this until the modified $M$ has full rank.

\begin{table}
    \caption{\label{tab:ccdbt-corrections}
    The six corrections needed to make the matrix $M$ invertible for the decomposition of $\ccnot{5}{3}{\mathcal{Z}}$ with $\ctrlseq$ from \Cref{eq:control-sequence}.
    Each correction is applied by conjugating the rotation gate corresponding to the (zero-indexed) column number of $M$ with a pair of GCX gates.
    In accordance with \Cref{eq:control-sequence}, we order the control register such that the auxiliary qutrit is first, followed by the four control link qutrits.
    For clarity, we adopt the qutrit labels in \Cref{fig:gggg-decomp}.
    }
    \begin{ruledtabular}
        \begin{tabular}{ccc}
        Column & Control qutrit & Control value \\
        \hline
        3  & $\ket{j^t_\ell}$ & $\ket{1}$ \\ %
        9  & $\ket{j^t_\ell}$ & $\ket{1}$ \\ %
        27 & $\ket{j^b_\ell}$ & $\ket{2}$ \\ %
        48 & $\ket{j^b_\ell}$ & $\ket{1}$ \\ %
        59 & $\ket{j^b_\ell}$ & $\ket{2}$ \\ %
        65 & $\ket{j^t_\ell}$ & $\ket{2}$    %
        \end{tabular}
    \end{ruledtabular}
\end{table}

For the $\ccnot{5}{3}{\mathcal{Z}}$ decomposition in \Cref{subsec:qutrit-plaquette-decomposition}, after applying the six corrections in \Cref{tab:ccdbt-corrections} to the original $M$, the modified matrix $\tilde{M}$ has full rank.
Since each correction consists of two GCX gates and two additional depth, 
the modified $\manccnot{5}{3}{\mathcal{Z}}$ decomposition uses 94 GCX gates, 82 $R_\mathcal{Z}$ gates, and has depth 176.

\section{Alternate $\uqtplaq$ Decomposition}
\label{app:gggg-decomp-alt}
\begin{figure}
    \includegraphics[width=\columnwidth]{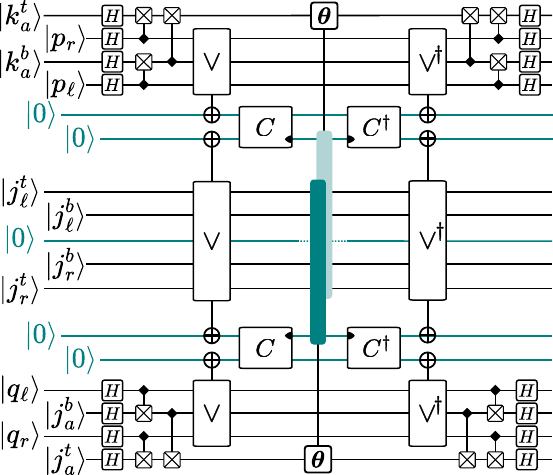}
    \caption{\label{fig:gggg-decomp-alt}
    Alternative decomposition of \Cref{fig:gggg-decomp} with {\FredkinTrick} applied to the control links in addition to the plaquette links.
    The input subspace information on the control links is computed, with the output copied onto two auxiliary qutrits.
    The $C$ gate (top of \Cref{fig:wide-qudit-lor-gates}) combines the input subspace information of the control links with the plaquette links to inform the gating of the $\ccnot{5}{3}{\mathcal{Z}}$ subroutine for opposite plaquette faces independently,
    maintaining the parallelized structure of the decomposition.
    The notch on the $C$ and $C^\dagger$ gates indicate the qutrit holding the combined input subspace information.
    The input size of the $\lor$ gates depend on the particular $\sumpi \GiTerm$ term (\Cref{tab:alt-decomp-lors}),
    with the circuit above showing the case for $\sumpi \Gi_{01}\Gi_{01}\Gi_{01}\Gi_{01}$, $\sumpi \Gi_{01}\Gi_{01}\Gi_{01}\Gi_{12}$, and $\sumpi \Gi_{01}\Gi_{12}\Gi_{01}\Gi_{12}$
    as well as their transformations under $D_4$ symmetry, covering $7$ of the $16$ $\sumpi \GiTerm$.
    }
\end{figure}

\noindent
In \Cref{fig:gggg-decomp-alt}, we present an alternative decomposition of $\uqtplaq$. 
Beginning from \Cref{fig:gggg-decomp}, we apply {\FredkinTrick} to the control links in addition to the plaquette links, enabling $\ccnot{5}{3}{\mathcal{Z}}$ decompositions with shorter control sequences.
The control sequences take advantage of the small size of set $F$ in \Cref{eq:control-set} in addition to the $(i+j+k+l)\bmod 2=0$ constraint.

For example, 
consider $\Gi_{01}\Gi_{01}\Gi_{01}\Gi_{01}$ (row 1 of \Cref{tab:qutrit-plaq-op}),
which applies controlled operations based on the control sequence
\begin{equation*}
    \{\ctrlseq\} = \{w \in \mathbb{Z}_2 \times \{v \in \mathbb{Z}_2^4 \enskip \mid v \bmod 2 = 0\}\},\ |\ctrlseq| = 16.
\end{equation*}
In this case, the control links are controlling only on a four qubit subspace.
By {\FredkinTrick} the control links, we can use the $\ccnot{5}{3}{\mathcal{Z}}$ subroutine 
with the $\ctrlseq$ above, which uses 16 GCX gates and 16 $R_\mathcal{Z}$ gates, instead of $\manccnot{5}{3}{\mathcal{Z}}$ with $\ctrlseq$ in \Cref{eq:control-sequence},
which uses 94 GCX gates and 82 $R_\mathcal{Z}$ gates.

\begin{table}
    \caption{\label{tab:alt-decomp-lors}
    The 16 $\sumpi \GiTerm$ separated into 4 groups by the input size of the applied $\lor$ gates.
    These $\lor$ gates are used in place of the four-input $\lor$ gates in \Cref{fig:gggg-decomp-alt}.
    The control links under {\FredkinTrick} permute in tandem with the $\GiTerm$ subspace and with the face of the cube applied (below, we apply to face A).
    }
    \begin{ruledtabular}
        \begin{tabular}{cccc}
        $\sumpi \GiTerm$ & order & $\lor$ gates & control links\\
        \hline
        $\Gi_{01}\Gi_{01}\Gi_{01}\Gi_{01}$ & 1 & Four-input & $\ket{j^t_\ell},\ket{j^b_\ell},\ket{j^b_r},\ket{j^t_r}$\\
        $\Gi_{01}\Gi_{01}\Gi_{01}\Gi_{12}$ & 4 & Four-input & $\ket{j^t_\ell},\ket{j^b_\ell},\ket{j^b_r},\ket{j^t_r}$\\
        $\Gi_{01}\Gi_{01}\Gi_{12}\Gi_{12}$ & 4 & Three-input &$\ket{j^t_\ell},\ket{j^b_\ell},\ket{j^b_r}$\\
        $\Gi_{01}\Gi_{12}\Gi_{01}\Gi_{12}$ & 2 & Four-input & $\ket{j^t_\ell},\ket{j^b_\ell},\ket{j^b_r},\ket{j^t_r}$\\
        $\Gi_{01}\Gi_{12}\Gi_{12}\Gi_{12}$ & 4 & Two-input  & $\ket{j^t_\ell},\ket{j^b_\ell}$\\
        $\Gi_{12}\Gi_{12}\Gi_{12}\Gi_{12}$ & 1 & No {\FredkinTrick}
        \end{tabular}
    \end{ruledtabular}
\end{table}
The set of control links covered by the gating subcircuit depends on the $\sumpi \GiTerm$ term.
In \Cref{tab:alt-decomp-lors}, we show the control links covered by the $\lor$ gates for each $\sumpi \GiTerm$ term.
The required $\lor$ gates are the two-input and three-input versions from \Cref{fig:qudit-lor-gates}, as well as the four-input $\lor$ gate in \Cref{fig:wide-qudit-lor-gates}.
We note that the four-input $\lor$ gate can be implemented with \Cref{fig:qudit-lor-gates} using a single auxiliary ququart.
For hardware implementations, it may be advantageous to invest in mixed-dimensional auxiliary qudits to reduce circuit width and depth.

We found that the $\ccnot{5}{3}{\mathcal{Z}}$ decompositions worked best (invertible $M$ and fewer gates) for $\exp(-it\sumpi \Gi_{pp^\prime}\Gi_{qq^\prime}\Gi_{rr^\prime}\Gi_{ss^\prime})$
where  $pqrs$ has a large integer representation.
Thus, we used decompositions for $pqrs \in \{0000, 1000, 1010, 1100, 1110\}$, permuting the order of the control qutrits to obtain the remaining $\sumpi \GiTerm$.
We use the modified $\manccnot{5}{3}{\mathcal{Z}}$ decomposition from Appendix~\ref{app:corr-82-ccdbt} to implement the ${pqrs = 1111}$ case.
We expect this dependence on ordering to be an artifact of the present implementation.

\begin{table}
    \caption{\label{tab:alt-decomp-gggg-resource}
    Resource count for two $\uqtplaq$applied in parallel to opposite faces of the cube.
    The $\exp(-it\sumpi \GiTerm)$ with $\lor$ gates of the same input size (\Cref{tab:alt-decomp-lors}) use the same resources.
    All $\lor$ gates on the control links use one additional GCX gate and depth to copy the output to another auxiliary qutrit.
    The total qutrit count is 17, where 12 are used to encode the gauge links and 5 are auxiliary.
    }
    \begin{ruledtabular}
        \begin{tabular}{c|ccccc|c}
    \multicolumn{7}{c}{$\sumpi \GiTerm$, no $\lor$ gates on control links}\\
        \hline
        Subcircuits & Counts & GCX & $R_\mathcal{Z}$ & $X$ & $H$ & Depth \\ 
        \hline
        $\manccnot{5}{3}{\mathcal{Z}}$ & $\times$ 2 & 94 & 82 & 0 & 0 & 176\\
        \hline
        \# $\sumpi \GiTerm$         & $\times$ 1 & 188 & 164 & 0 & 0 & 176\\
        \hline
        Total Contribution          &            & 188 & 164 & 0 & 0 & 176\\
        \hline \hline
    \multicolumn{7}{c}{$\sumpi \GiTerm$, two-input $\lor$ on control links}\\
        \hline
        Subcircuits & Counts & GCX & $R_\mathcal{Z}$ & $X$ & $H$ & Depth \\ 
        \hline
        $\ccnot{5}{3}{\mathcal{Z}}$ & $\times$ 2 & 36 & 36 & 0 & 0 & 72\\
        Control $\lor$              & $\times$ 2 & 4 & 0 & 0 & 0 & 4$\times$2\\
        \hline
        \# $\sumpi \GiTerm$         & $\times$ 4 & 80 & 72 & 0 & 0 & 80\\
        \hline
        Total Contribution          &            & 320 & 288 & 0 & 0 & 320\\
        \hline \hline
        \multicolumn{7}{c}{$\sumpi \GiTerm$, three-input $\lor$ on control links}\\
        \hline
        Subcircuits & Counts & GCX & $R_\mathcal{Z}$ & $X$ & $H$ & Depth \\ 
        \hline
        $\ccnot{5}{3}{\mathcal{Z}}$ & $\times$ 2 & 24 & 24 & 0 & 0 & 48\\
        Control $\lor$              & $\times$ 2 & 7 & 0 & 1 & 0 & 7$\times$2\\
        \hline
        \# $\sumpi \GiTerm$         & $\times$ 4 & 62 & 48 & 2 & 0 & 62\\
        \hline
        Total Contribution          &            & 248 & 192 & 8 & 0 & 248\\
        \hline \hline
    \multicolumn{7}{c}{$\sumpi \GiTerm$, four-input $\lor$ on control links}\\
        \hline
        Subcircuits & Counts & GCX & $R_\mathcal{Z}$ & $X$ & $H$ & Depth \\ 
        \hline
        $\ccnot{5}{3}{\mathcal{Z}}$ & $\times$ 2 & 16 & 16 & 0 & 0 & 32\\
        Control $\lor$              & $\times$ 2 & 9 & 0 & 1 & 0 & 7$\times$2\\
        \hline
        \# $\sumpi \GiTerm$         & $\times$ 7 & 50 & 32 & 2 & 0 & 46\\
        \hline
        Total Contribution          &            & 350 & 224 & 14 & 0 & 322\\
        \hline \hline
    \multicolumn{7}{c}{Common Subcircuits}\\
        \hline
        Subcircuits & Counts     & GCX & $R_\mathcal{Z}$ & $X$ & $H$ & Depth \\ 
        \hline
        $\mathcal{X}$-parity       & $\times 4 \times 16$ & 3 & 0 & 0 & 4 & 3$\times$2\\
        $C$ gate        & $\times 4 \times 15$ & 2 & 0 & 1 & 0 & 3$\times$2\\
        Plaquette $\lor$ & $\times 4 \times 16$ & 6 & 0 & 1 & 0 & 6$\times$2\\
        \hline
        Total Contribution          &            & 696 & 0 & 124 & 256 & 378\\
        \hline \hline
        Total (two $\qtplaq$)       &            & 1802 & 868 & 146 & 256 & 1444\\
        \end{tabular}
    \end{ruledtabular}
\end{table}

\Cref{tab:alt-decomp-gggg-resource} shows the resource count for two $\uqtplaq$ applied in parallel to opposite faces of the cube.
Compared to the first row of \Cref{tab:cube-sim-resource}, we save $3584-1802=1782$ GCX gates and depth is reduced to $3104-1444=1660$,
which is around half the resources at the cost of three additional auxiliary qutrits (on a purely qutrit platform).

\section{Simulation Results}
\label{app:simulation-results}
\noindent
The global basis used to obtain the exact simulation results in \Cref{fig:results} 
(and shown in \Cref{tab:simulation-exact}) comprises only physical states of the cube.
\begin{table}
    \caption{\label{tab:simulation-exact}
    Exact simulation using the global basis of physical states on a cube.
    It took %
    16 seconds to generate this table.
    }
    \begin{ruledtabular}
        \begin{tabular}{cc}
        Time ($t$) & $\langle \hat{H}_{E,\square_A} \rangle$\\
        \hline
        0.02 & 0.0059995205375 \\
        0.12 & 0.2013777323202 \\
        0.22 & 0.4735643212012 \\
        0.32 & 0.5530191097325 \\
        0.42 & 0.4390914030334 \\
        0.52 & 0.2780532551655 \\
        0.62 & 0.2269820889009 \\
        0.72 & 0.3062084922296 \\
        0.82 & 0.4407325362572 \\
        0.92 & 0.4749953963899 \\
        \end{tabular}
    \end{ruledtabular}
\end{table}
The observable is the expectation value of the electric operator of plaquette $A$
\begin{equation}
    \hat{H}_{E,\square_A} = \frac{g^2}{2} \sum_{\text{links} \in \square_A} \sum_{j=0}^{\Lambda_j} j(j+1) \Pi_{2j}.
    \label{eq:exp-val-elec}
\end{equation}

We report the expectation values of the cube simulation for Trotter steps $1$ and $2$ in 
\begin{table}
    \caption{\label{tab:simulation-trotterized}
    The measured observable is the expectation value of $\hat{H}_E$ for face A, shown in \Cref{fig:results}.
    The runtime of the first Trotter step is significantly longer due to overhead incurred by just-in-time (JIT) compilation, performed by JAX.
    After that, the runtime scales linearly with the number of Trotter steps $N_T$.
    As an example, the average runtime for $N_T=7$ is 278 seconds.
    We use a time step of 0.05 but show every second expectation value for conciseness.
    }
    \begin{ruledtabular}
        \begin{tabular}{ccc}
        Time ($t$) & $\langle \hat{H}_{E,\square_A} \rangle$ & Runtime (seconds)\\
        \hline \hline
        \multicolumn{3}{c}{$N_T=1$} \\
        \hline
        0.02 & 0.00607722 & 367.325 \\
        0.12 & 0.23791493 &  42.091 \\
        0.22 & 0.57592832 &  42.295 \\
        0.32 & 0.74714905 &  42.006 \\
        0.42 & 0.58758336 &  42.043 \\
        0.52 & 0.34950834 &  42.102 \\
        0.62 & 0.00206689 &  42.068 \\
        0.72 & 0.16572029 &  42.031 \\
        0.82 & 0.44481068 &  42.074 \\
        0.92 & 0.67250186 &  42.310 \\
        \hline \hline
        \multicolumn{3}{c}{$N_T=2$} \\
        \hline
        0.02 & 0.00603121 & 80.451 \\
        0.12 & 0.21797391 & 80.381 \\
        0.22 & 0.48899010 & 80.128 \\
        0.32 & 0.56093830 & 80.621 \\
        0.42 & 0.52352385 & 84.377 \\
        0.52 & 0.46890193 & 84.309 \\
        0.62 & 0.43751637 & 78.712 \\
        0.72 & 0.42842750 & 78.745 \\
        0.82 & 0.40681914 & 78.721 \\
        0.92 & 0.41583451 & 78.728 \\
        \end{tabular}
    \end{ruledtabular}
\end{table}
\Cref{tab:simulation-trotterized}.
We note the first timestep takes significantly longer to execute. 
This is due to the use of just-in-time (JIT) compilation with JAX, which is applied to multiple
subroutines in the cube simulation. The first time a JIT-compiled function executes,
a compiled binary is generated and cached for future use, yielding a significant speed
boost.

\section{\label{app:hadamard-gcx}Hadamard Gate Removal}
\noindent
In the decomposition of the plaquette operator unitary evolution, 
the subspaces of $H$ and GCX gates originate from the subspace of $\Gi$, which are neighboring basis states in the SU(2) gauge theory.
\begin{figure}
    \[
        \Qcircuit @C=.5em @R=.5em {
        & \gate{H_{01}} & \ctrl{1} & \gate{H_{01}} & \qw \\
        & \gate{H_{01}} & \gate{X_{01}} & \gate{H_{01}} & \qw 
        } =
        \Qcircuit @C=.5em @R=1.2em {
        & \gate{X_{01}} & \qw \\
        & \ctrl{-1}     & \qw 
    } 
    \]
    \[
        \Qcircuit @C=.5em @R=.5em {
        & \gate{H_{01}} & \ctrl{1} & \gate{H_{01}} & \qw \\
        & \gate{H_{12}} & \gate{X_{12}} & \gate{H_{12}} & \qw 
        } =
        \Qcircuit @C=.5em @R=.5em {
        & \gate{X_{01}} & \qw  \\
        & \tctrlt{-1}   & \qw  
        } 
    \]
    \[
        \Qcircuit @C=.5em @R=.5em {
        & \gate{H_{12}} & \tctrlt{1} & \gate{H_{12}} & \qw \\
        & \gate{H_{01}} & \gate{X_{01}} & \gate{H_{01}} & \qw 
        } =
        \Qcircuit @C=.5em @R=1.2em {
        & \gate{X_{12}} & \qw  \\
        & \tctrlo{-1}    & \qw  
        } 
    \]
    \[
        \Qcircuit @C=.5em @R=.5em {
        & \gate{H_{12}} & \tctrlt{1} & \gate{H_{12}} & \qw \\
        & \gate{H_{12}} & \gate{X_{12}} & \gate{H_{12}} & \qw 
        } =
        \Qcircuit @C=.5em @R=1.2em {
        & \gate{X_{12}} & \qw  \\
        & \tctrlt{-1}    & \qw  
        } 
    \]
    \caption{\label{fig:hadamard-flip-gcx}
        All $H$ gates can be removed by flipping the GCX gate and changing the control value and $X_{ij}$ subspace.
    }
\end{figure}
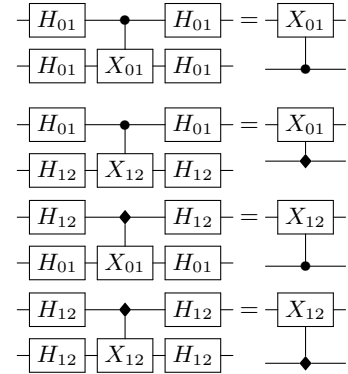
Thus, for a qutrit digitization of the gauge field, only configurations in \Cref{fig:hadamard-flip-gcx} appear.
As can be seen in \Cref{fig:qutrit-circuit}, the $H$ gates on the left can be propagated through the circuit until they cancel with those on the right,
transforming $R_{\mathcal{Z}} \rightarrow R_{\mathcal{X}}$ and GCX $\rightarrow$ GCZ along the way.
Therefore, we can remove all $H$ gates from the circuit, and by extension, the entire cube simulation.

\FloatBarrier

\bibliography{
  bibs/qlgt-repr,
  bibs/qudit-hardware,
  bibs/qudit-simulations,
  bibs/qudit-compilation,
  bibs/qudit-computing,
  bibs/misc
}

\end{document}